\documentclass[aps,prapplied,showpacs,twocolumn,preprintnumbers,amsmath,amssymb,longbibliography]{revtex4-2}
\usepackage[ruled,lined,boxed,commentsnumbered]{algorithm2e}
\usepackage{amsbsy}
\usepackage{amsfonts}
\usepackage{amssymb}
\usepackage{bm}

\usepackage{graphicx}% Include figure files
\usepackage{dcolumn}% Align table columns on decimal point
\usepackage{bm}% bold math
\usepackage{color}
\usepackage{amsmath}
\usepackage{upgreek}
\usepackage{float}
\usepackage[colorlinks,linkcolor=blue, urlcolor=blue, anchorcolor=blue, citecolor=blue]{hyperref}

\begin{document}

\title{Photonic restricted Boltzmann machine for content generation tasks }% Force line breaks with \\
\author{ Li Luo }
\thanks{These authors contributed equally to this work.}
\affiliation{ School of Physics, State Key Laboratory of Extreme Photonics and Instrumentation, and Zhejiang Province Key Laboratory of Quantum Technology and Device, Zhejiang University, Hangzhou 310027, China}

\author{ Yisheng Fang }
\thanks{These authors contributed equally to this work.}
\affiliation{ School of Physics, State Key Laboratory of Extreme Photonics and Instrumentation, and Zhejiang Province Key Laboratory of Quantum Technology and Device, Zhejiang University, Hangzhou 310027, China}

\author{ Wanyi Zhang}
\affiliation{ School of Physics, State Key Laboratory of Extreme Photonics and Instrumentation, and Zhejiang Province Key Laboratory of Quantum Technology and Device, Zhejiang University, Hangzhou 310027, China}

\author{Zhichao Ruan}
\email{zhichao@zju.edu.cn}
\affiliation{ School of Physics, State Key Laboratory of Extreme Photonics and Instrumentation, and Zhejiang Province Key Laboratory of Quantum Technology and Device, Zhejiang University, Hangzhou 310027, China}

\begin{abstract}
The restricted Boltzmann machine (RBM) is a neural network based on the Ising model, well known for its ability to learn probability distributions and stochastically generate new content. However, the high computational cost of Gibbs sampling in content generation tasks imposes significant bottlenecks on electronic implementations. Here, we propose a photonic restricted Boltzmann machine (PRBM) that leverages photonic computing to accelerate Gibbs sampling, enabling efficient content generation. By introducing an efficient encoding method, the PRBM eliminates the need for computationally intensive matrix decomposition and reduces the computational complexity of Gibbs sampling from $O(N)$ to $O(1)$. Moreover, its non-Von Neumann photonic computing architecture circumvents the memory storage of interaction matrices, providing substantial advantages for large-scale RBMs. We experimentally validate the photonic-accelerated Gibbs sampling by simulating a two-dimensional Ising model, where the observed phase transition temperature closely matches the theoretical predictions. Beyond physics-inspired tasks, the PRBM demonstrates robust capabilities in generating and restoring diverse content, including images and temporal sequences, even in the presence of noise and aberrations. The scalability and reduced training cost of the PRBM framework underscore its potential as a promising pathway for advancing photonic computing in generative artificial intelligence.
\end{abstract}
\maketitle

\section{Introduction}

Restricted Boltzmann machine (RBM) is a stochastic generative neural network based on the Ising model, characterized by spin interactions solely between the visible and hidden layers \cite{smolensky1986information,hinton2002training}. RBMs can be applied to a variety of tasks by learning the probability distribution of the training data. It is renowned for the development of deep learning \cite{salakhutdinov2009deep,hinton2006reducing,bojnordi2016memristive,goodfellow2016deep} and for demonstrating features in classification and recognition tasks, particularly generating new content \cite{goodfellow2016deep, larochelle2008classification,  fernandez2023disentangling,patel2022logically, li2024restricted}. Furthermore, by combining RBM with recurrent neural network, the time-varying RBM enables effective modeling of temporal data \cite{boulanger2012modeling}. In RBMs, Gibbs sampling plays a crucial role by decomposing high-dimensional problems into multiple conditional distribution problems with Markov chains, and sampling sequentially from the conditional distribution \cite{koller2009probabilistic,andrieu2003introduction}. However, traditional electronic computing struggles with long chains of Gibbs sampling, especially when dealing with large-scale datasets, which demand enormous computational resources \cite{goodfellow2016deep, ackley1985learning, long2010restricted, niazi2024training}.

Recently, there has been growing interest in developing unconventional computational architectures for simulating Ising Hamiltonians to accelerate specific computations. For example, to solve NP-hard problems, various Ising machines have been developed based on optical parametric oscillators \cite{mcmahon2016fully,inagaki2016coherent,inagaki2016large,bohm2019poor,marandi2014network}, lasers \cite{utsunomiya2011mapping, babaeian2019single, tradonsky2019rapid,parto2020realizing, honari2020mapping}, polariton \cite{kalinin2020polaritonic, berloff2017realizing, kalinin2018simulating},  trapped ions \cite{kim2010quantum}, atomic and photonic condensates \cite{struck2013engineering, vretenar2021controllable}, electronic memorisers \cite{cai2020power}, superconducting qubits \cite{johnson2011quantum, boixo2014evidence, king2018observation}, and nanophotonics circuits \cite{roques2020heuristic,prabhu2020accelerating,shen2017deep,wu2014optical, okawachi2020demonstration,prabhu2020accelerating,hua2025integrated, wu2025monolithically,jiang2025photonic,ouyang202516}. In particular, the spatial photonic Ising machine (SPIM) has been demonstrated with reliable large-scale Ising spin systems, even up to ten-thousands of spins \cite{pierangeli2019large}. By leveraging multiple degrees of freedom of light, such as time \cite{yamashita2023low}, space \cite{veraldi2025fully,shimomura2025parallel, sakellariou2025encoding}, and wavelength \cite{luo2023wavelength}, several division multiplexing schemes have been developed, harnessing the advantages of the ultrafast-speed and high-power-efficiency photonic computation \cite{yamashita2023low, luo2023wavelength,veraldi2025fully, shimomura2025parallel,sakellariou2025encoding}. Unlike traditional Von Neumann architectures, SPIMs encodes both the spin interaction matrix and spin configurations using spatial light modulators (SLMs) \cite{pierangeli2019large,fang2021experimental,pierangeli2020noise,huang2021antiferromagnetic,yamashita2023low,sun2022quadrature,kumar2023observation} or digital micromirror devices (DMDs) \cite{leonetti2021optical, yu2024spatial}, with measuring light intensity and instantaneously computing Ising Hamiltonian with Mattis-type interaction \cite{pierangeli2019large, leonetti2021optical,fang2021experimental, pierangeli2020noise}, antiferromagnetic interaction \cite{huang2021antiferromagnetic}, low-rank interaction \cite{yamashita2023low, sun2022quadrature}, four-body interaction \cite{kumar2023observation}. However, all these Ising machines are designed to simulate Ising model wherein spins reside within a single layer and interact with each other, whereas RBMs partition spins into distinct visible and hidden layers, with spin interactions existing solely between two layers. Therefore, these computation schemes are unable to accelerate the computation for RBMs, due to the discrepancy of the spin distribution and interaction.

\begin{figure*}
\centerline{\includegraphics[width=5.6in] {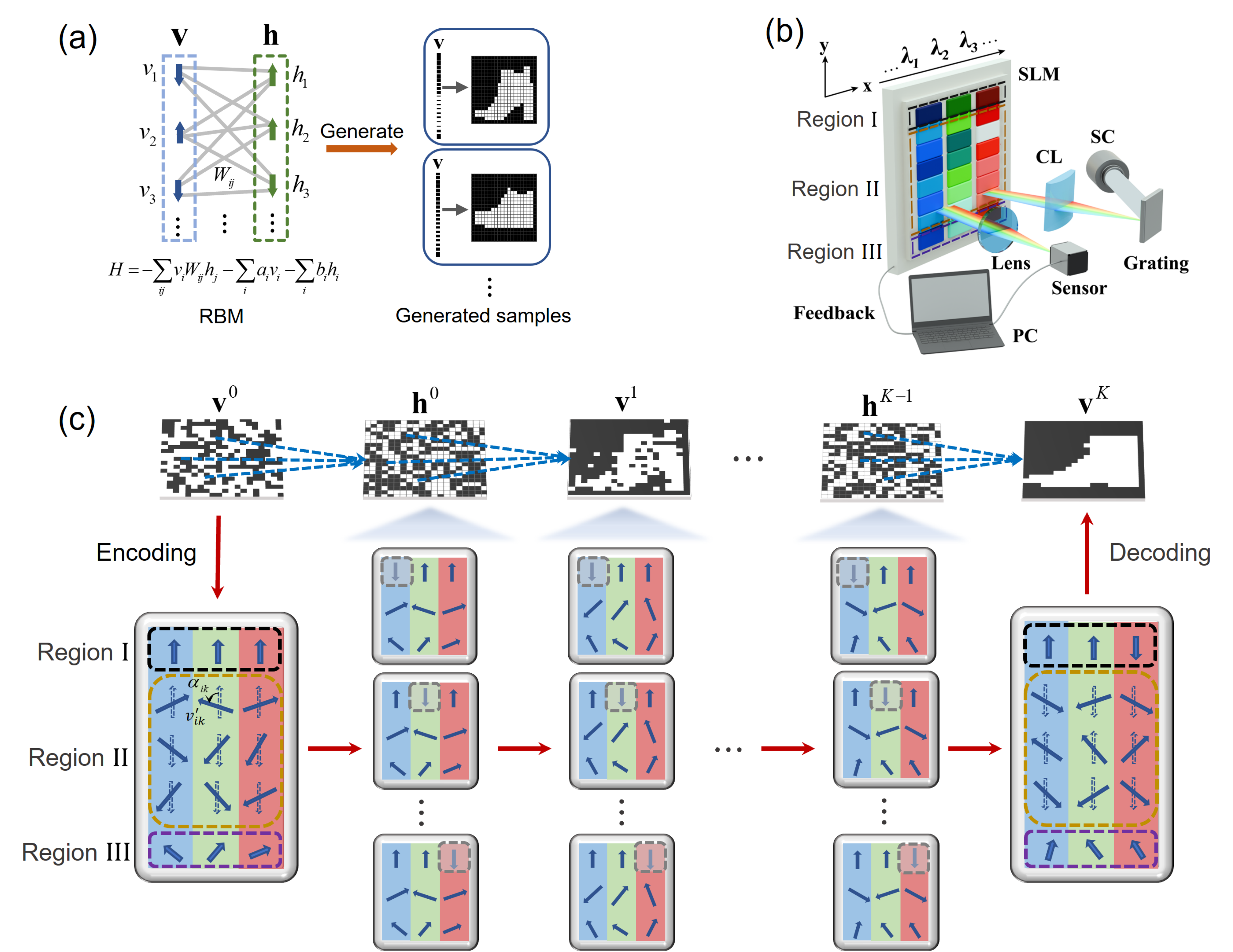}}
\caption{\label{fig:1}\textbf{ Schematic of photonic restricted Boltzmann machine with Gibbs sampling.} (a)The RBM consists of visible and hidden layers of spins ${\bf{v}}$ and ${\bf{ h}}$, with spin interactions $ W_{ij}$ solely between two layers and under external magnetic fields ${a_i}$ and ${b_j}$, respectively. For content generation tasks, RBM is trained by maximizing the log-likelihood function and performed to generate new contents, by Gibbs sampling. (b) Schematic of the PRBM architecture. In this setup, light with different wavelengths is diffracted and focuses on a phase-only spatial light modulator (SLM) along the $x$-direction, while the pixels in the $y$-direction are coherently illuminated by incident light of the same wavelength. The spins are encoded with phase modulation on the SLM at three different regions (circuited by dashed lines). SC, supercontinuum laser; CL, cylindrical lens; Sensor, sCMOS camera. (c) Experimental process of content generation by PRBM. Initial random visible spin configuration is encoded onto Region\;{\rm II}. By performing iterative photonic Gibbs sampling between the visible and hidden layers, a generated content is decoded from the final visible layer.}
\end{figure*}

In this article, we propose a photonic restricted Boltzmann machine (PRBM) that leverages photonic computing to accelerate Gibbs sampling and demonstrate content generation tasks. The PRBM is designed based on wavelength-division multiplexing spatial Ising machines, inheriting the scalability of SPIM architecture. By developing an encoding method, we eliminate the need for the decomposition computation of interaction matrix and reduce the computational complexity of Gibbs sampling from $O(N)$ to $O(1)$, thereby significantly reducing computational costs. In addition, with the advantage of photonic computation of non-Von Neumann computing architecture, we circumvent the memory storage of interaction matrix, substantially enhancing computation efficiency for large-scale RBM. We further validate Gibbs sampling process by simulating analytically solvable two-dimensional Ising model \cite{onsager1944crystal} and experimentally observe the phase transition at the critical temperature agreeing well with exact theoretical value. Moreover, we experimentally demonstrate that the proposed PRBM effectively acquires the capability of generating new content, with successful implementations in image generation and temporal data processing, as exemplified by music generation. Finally, we prospect the scalability and the training advances of the PRBM framework, suggesting its potential to offer a promising pathway for applications of generative artificial intelligence (AI).

\section{Photonic computing scheme of restricted Boltzmann machine}
We consider RBMs with a Hamiltonian $H({\bf{v,h}}) =  - \sum\limits_{ij} {{v_i}{W_{ij}}{h_j} - \sum\limits_i {{a_i}{v_i}}  - } \sum\limits_i {{b_i}{h_i}}$, where ${v_i}$ and ${h_i}$ taking values of $ \pm 1$ represent the spin configurations in the visible and the hidden layers. The spins interact between the two layers solely as $W_{ij}$, where ${a_i}$ and ${b_i}$ are the magnetic field terms for visible layer spins and hidden layer spins, respectively (see Figure 1(a) and the Methods). We note that SPIMs simulating such spin interactions in RBMs has to introduce substantial redundant zero-value elements \cite{bohm2022noise} and further requires digital computation of eigenvalue decomposition \cite{yamashita2023low, veraldi2025fully} or Cholesky-like decomposition \cite{luo2023wavelength} with the cost of computation complexity $O({N^3})$. Such computation cost becomes significant when the total spin number in the visible and hidden layers is large, and is further intensified when accounting for external magnetic fields on each spin. Additionally, SPIMs utilize Metropolis-Hastings (MH) sampling evaluating the impact of a single spin flip on the global system, whereas Gibbs sampling evaluates the impact of the locally correlated system on a single spin, which is crucial for accelerating computation in complex high-dimensional problems. Therefore, how to effectively encode the visible and hidden layers and how to implement Gibbs sampling constitute two key challenges in demonstrating the photonic computing acceleration for RBMs.

Figure 1(b) shows the proposed PRBM by the photonic computing of Gibbs sampling, following the measurement and feedback scheme. Here a collimated supercontinuum laser is diffracted by a grating and different wavelength light focuses on a SLM along the $x$ axis with a cylindrical lens. The modulated light beam reflected from SLM is transformed by a lens, resulting in an incoherent intensity summation of different wavelengths and coherent interference for each wavelength at the back focal plane (detailed information is provided in Supplementary Materials Section $\rm I$ ).

To enable the photonic computing of Gibbs sampling, we divide the SLM into three distinct regions along the $y$ axis. We take the example of the generation process as illustrated in Figure 1(c). When sampling from the visible layer ${\mathbf{v}}$ to the hidden layer ${\mathbf{h}}$, for each hidden-layer spin ${h_k}$, the visible-layer spin ${{v_i}}$ with the spin interaction ${W_{ik}}$ and under the magnetic field ${b_k}$ are encoded at the $k$-th wavelength, and the phase modulation along the $y$ axis of the SLM are given as
\begin{equation}
\begin{array} {ll}
\text{Region}\;{\rm I}:  &\{ \phi _{k}^{m,n}\} ^{\rm I} = {h_k}\frac{\pi }{2} \\
\text{Region}\;{\rm II}:  &\{ \phi _{i,k}^{m,n}\} ^{{\rm II}} = {v_i}\frac{\pi }{2} + {( - 1)^{m + n}}{\alpha _{ik}} \\
\text{Region}\;{\rm III}: &\{ \phi _{k}^{m,n}\} ^{{\rm III}} = s\frac{\pi }{2} + {( - 1)^{m + n}}{\beta _k}.  \label{eq:1}
\end{array}
\end{equation}
Here an auxiliary spin $s = 1$ is used for magnetic field ${b_k}$ in Region ${\rm III}$. We encode each spin by a macro-pixel which consists of ${N_x} \times {N_y}$ pixels, where $m$ and $n$ are the pixel indices within each spin, $1 \leqslant m \leqslant {N_x}$ and $1 \leqslant n \leqslant {N_y}$.

In order to implement the spin interaction and the external magnetic field, we propose the Gauge transform \cite{fang2021experimental, huang2021antiferromagnetic}: In Region ${\rm II}$ and ${\rm III}$, the phase modulations have additional contributions from a checkerboard modulation, corresponding to rotating each spin by the angle ${\alpha _{ik}} = \arccos ({W_{ik}}/L)$ and ${\beta _k} = \arccos ({b_k}/L)$, which are equivalent spins $v_{ik}^{'} = {v_i}{W_{ik}}$ and $s_k^{'} = s{b_k}$, respectively. The normalization parameter $L$ is the maximum absolute value of the interactions and external magnetic fields.

With this encoding method, we perform Gibbs sampling in two steps. In the first step, we set all hidden layer spins as ${h_i} = 1$ and encode them with visible layer spins onto the SLM following Equation~\ref{eq:1}. Thus, the normalized intensity ${\tilde I_0}$ at the center position of the back focal plane is obtained, and the Hamiltonian is computed as ${H_0} =  - L{\tilde I_0} =  - 2\sum\limits_{i = 1}^{{N_h}} {\sum\limits_{j = 1}^{{N_v}} {{v_j}{W_{ji}} - } } 2\sum\limits_{i = 1}^{{N_h}} {{b_i} - } Q$, where $Q$ is a constant, ${N_v}$ and ${N_h}$ are the number of the spins in the visible and hidden layers, respectively (Supplementary Materials Section $\rm {II}$). In the second step, every spin in the hidden layer is flipped independently, as indicated by the gray shadowed box in Figure 1(c). For the $k$-th spin, we set ${h_k} =  - 1$ and measure the intensity corresponding to the Hamiltonian ${H_k} =  - 2\sum\limits_{i \ne k}^{{N_h}} {\sum\limits_{j = 1}^{{N_v}} {{v_j}{W_{ji}} - } } 2\sum\limits_{i \ne k}^{{N_h}} {{b_i} + 2\sum\limits_{j = 1}^{{N_v}} {{v_j}{W_{jk}}}  + 2{b_k} - Q}$. Therefore, by measuring the difference of intensity in two steps, we can compute $\Delta H_k $ as
\begin{equation}
\Delta H_k = ({H_0} - {H_k})/2 =  - 2\sum\limits_{j = 1}^{{N_v}} {{v_j}{W_{jk}}}  - 2{b_k}. \label{eq:2}
\end{equation}
Based on the feedback scheme, we update the spin configuration with Gibbs sampling at temperature $T$, by calculating the probability of the $k$-th spin being $+1$ as $P({h_k} = 1|{\bf{v}}) = 1/[1 + \exp (  \Delta H_k/T)]$.

We note that the proposed encoding method for Gibbs sampling does not require the decomposition of the spin interaction matrix \cite{yamashita2023low, veraldi2025fully, luo2023wavelength} and can be generalized for sparse spin interactions (see more details in the Methods). When sampling from the hidden layer to the visible layer, we encode the visible layer spin ${v_k}$ and the external magnetic field ${a_k}$  in region ${\rm I}$ and ${\rm III}$, respectively, and the equivalent spin $h_{ik}^{'} = {h_i}{W_{ki}}$ in Region ${\rm II}$. In the process of generating an image, we begin with a randomized visible layer spin configuration encoded onto Region\;{\rm II}. By performing iterative photonic computing of Gibbs sampling between the visible and hidden layers, a generated content is decoded from the final visible layer as shown in Figure 1(c).

\begin{figure}
\centerline{\includegraphics[width=3.2in]{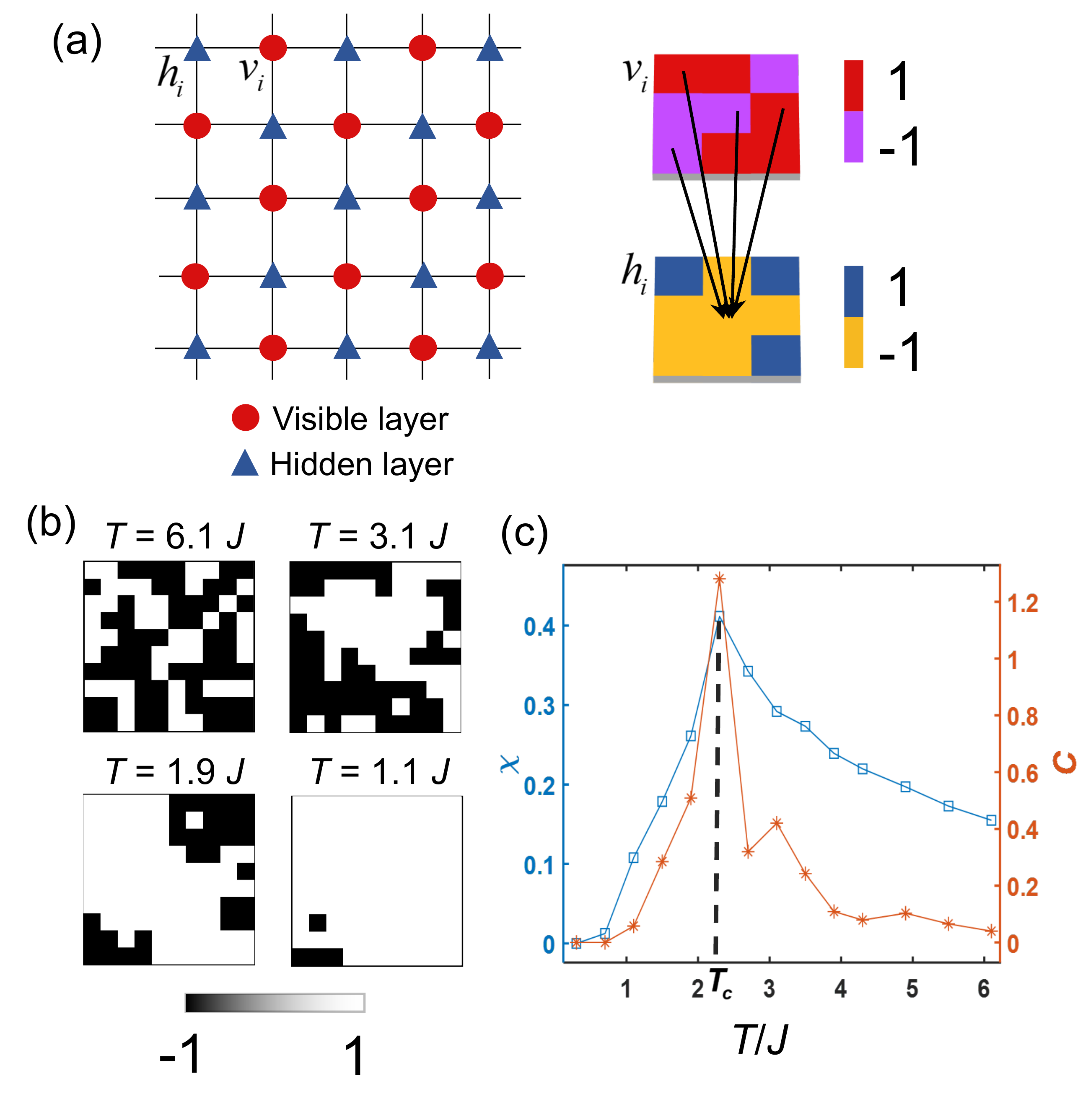}}
\caption{\label{fig:2}\textbf{Validation of PRBM by observing phase transition in two-dimensional Ising model.} (a) Left: The spins in the 2D Ising model are divided into visible and hidden layers, with each spin interacting only with its nearest-neighboring spins. Right: The visible layer spins are color-coded as red and purple, while blue and yellow correspond to spins in the hidden layer. (b) The equilibrium states sampling from four different temperatures by Gibbs sampling through photonic computing. (c) The variation of the order parameter susceptibility and heat capacity with temperature, and the phase transition temperature is $T_c=2.3J$.}
\end{figure}

\section{Validation of PRBM by observing phase transition}
We first validate the proposed PRBM with Gibbs sampling by observing the phase transition in the two-dimensional Ising model $H=-J\sum\limits_{<ij>}\sigma_i \sigma_j $, where each spin interacts only with its four nearest neighbors of the strength $J$. Analogous to RBM, we transform the spins at the four corner positions of the square into visible layer (red circular dots in Fig. 2(a)) and hidden layer (blue triangles in Fig. 2(a)), respectively. Through this transformation, we study the phase transition and compare the experimental results with the analytical critical temperature \cite{onsager1944crystal}. We experimentally conduct photonic computing of the Gibbs sampling on a $10 \times 10$ lattice at fourteen different temperatures. In order to calculate the phase diagram, we generate 20 samples at each temperature, with each sample obtained after 10 iterations from a random initial spin configuration, with each complete iteration consisting of Gibbs sampling from the visible layer to obtain the hidden layer, and vice versa. Figure 2(b) displays the obtained spin configurations after 10 iterations at four different temperatures $T = 6.1J,3.1J,1.9J$, and $1.1J$. At high temperature, such as $T = 6.1J$, the spin configuration exhibits disordered, while as the temperature decreases, cluster of aligned spins begin to emerge and the spin configuration becomes increasingly ordered.

To precisely evaluate the phase transition temperature, we calculate the heat capacity $C = \frac{{\left\langle {{H^2}} \right\rangle  - {{\left\langle H \right\rangle }^2}}}{{N{T^2}}}$ and susceptibility $\chi  = \sum\limits_{i = 1}^N {\frac{{1 - m_i^2}}{{NT}}}$, where ${m_i}$ represents the statistical average of a spin and $N$ is the number of spins. As shown in Figure 2(c), the experimental results show that the phase transition temperature is ${T_c} = 2.3J$, which agrees well with the theoretical predictions ${T_c} = \frac{{2J}}{{\ln (1 + \sqrt 2 )}} \approx 2.27J$ in the infinite 2D model \cite{onsager1944crystal}. We also employ the annealing method \cite{kirkpatrick1983optimization} with Gibbs sampling to search the ground state (Supplementary Materials Section $\rm III$). All these results demonstrate that, despite limited noise and aberrations, the proposed photonic computing of Gibbs sampling effectively simulates the spin interaction between the visible and hidden layers for RBMs.

\begin{figure}
\centerline{\includegraphics[width=3.0in]{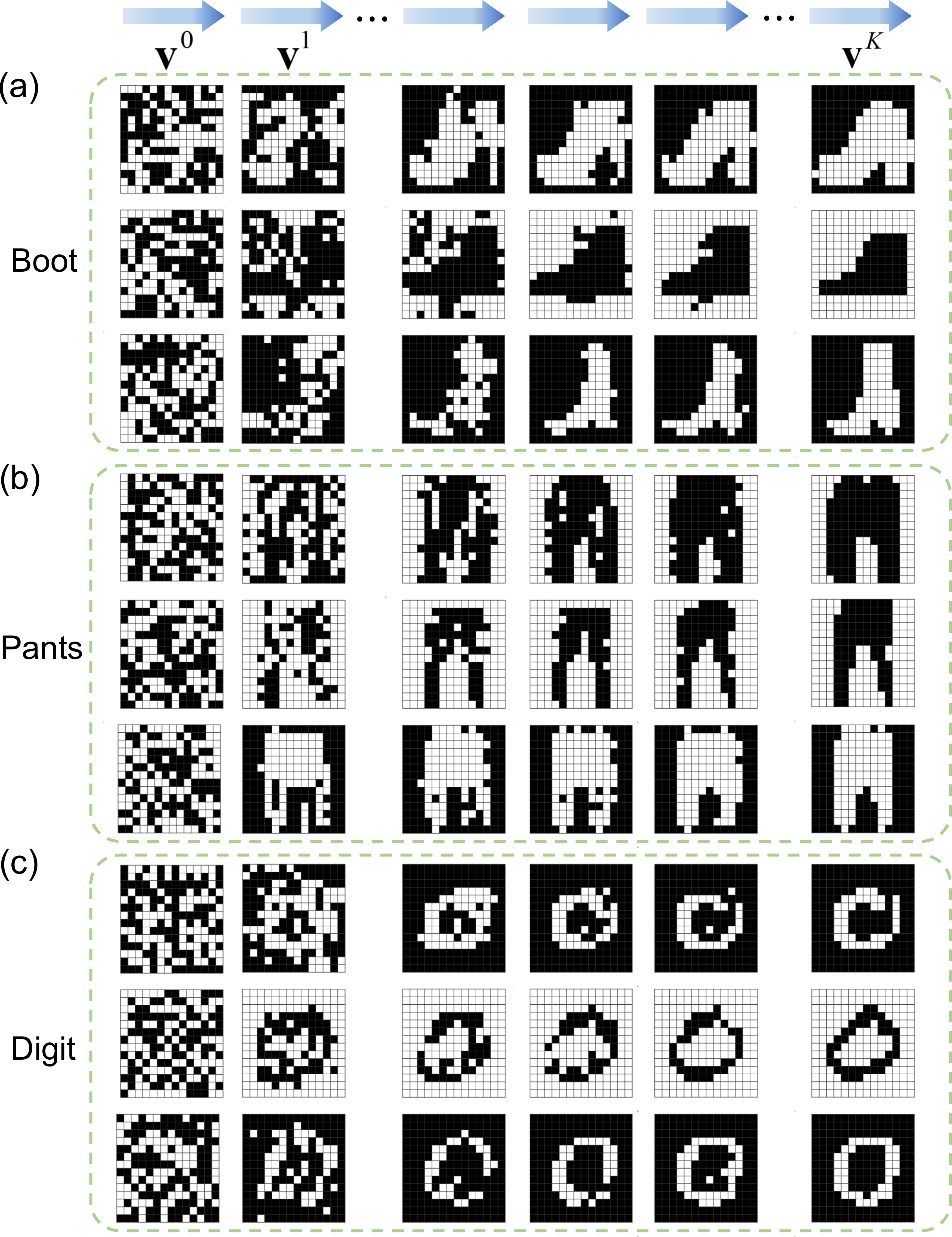}}
\caption{\label{fig:3}\textbf{Image content generation with PRBM.} (a)-(c) The processes of generating three independent images for each of `Boot', `Pants', and the digit `0', respectively.}
\end{figure}

\section{Image content generation and restoration by PRBM}
Next we demonstrate the capability of the photonic restricted Boltzmann machine for content generation. In image generation, we utilize the Fashion MNISIT dataset \cite{xiao2017/online} and MNIST digit images \cite{YannThe} as the visible layer spins to train the RBM, where each pixel in an image corresponds to a spin. Before training, we compress the original images with $28 \times 28$ spins to a resolution of $14 \times 14$ (yielding 196 spins) and binarize the images. During both the training and generation processes, Gibbs sampling is conducted at a fixed temperature $T = 1$. Furthermore, we train the model in the absence of magnetic field and separately train it on the fashion items `Boot', `Pants', and the digit `0'. Following the Hinton recipe \cite{hinton2012practical}, we initialize the interaction weights with small random values drawn from a zero-mean Gaussian distribution (Supplementary Materials Section $\rm {IV}$).

To generate an image by PRBM, we first randomly initialize the state vector ${{\mathbf{v}}^0}$ and subsequently obtain ${{\mathbf{v}}^K}$ as the terminal generated image after $K = 15$ iterations by performing photonic computing of Gibbs sampling. With the stabilized spin configuration in the visible layer, Figure 3 illustrates three distinct image generation processes for each experimental item. Additional experimental results on image generation are provided in Supplementary Materials Section $\rm {VII}$. The diversity observed across these generated images demonstrates the PRBM's capacity to produce new content with significant variability.

We further validate photonic computing of Gibbs sampling by evaluating whether the trained model exhibits overfitting. Here we use PRBM to perform restoring-image test, where the initial images are excluded in the training dataset and distorted. We compress images of the fashion item `Boot' and the digit `0' to $20 \times 20$ spins and train the model. For the image restoration experiments, we mask three different `Boot' images randomly selected from the test dataset, as indicated by the red areas in Figure 4(a). Figure 4(b) represents the initial input images ${{\mathbf{v}}^0}$ (the first column), where spins in the masked regions are set to a value of $-1$. After $K = 15$ iterations, the three restored images ${{\mathbf{v}}^K}$ are obtained, as shown in the last column of Figure 4(b).

As another example, we also restore three different `0' images from the test dataset by adding random noise, as indicated by the red marks in Figure 4(c). Before performing image restoration, we set the spins in the red-marked regions to the negative values of their respective original values, as is shown in the first column ${{\mathbf{v}}^0}$ of Figure 4(d). After 15 iterations, the final restored images ${{\mathbf{v}}^K}$ are presented in the last column of Figure 4(d). More other experimental results about image restoration are shown in Supplementary Materials Section $\rm {VII}$). All these results demonstrate that the image generation capabilities of the PRBM are not attributable to overfitting and also validate the proposed photonic computing of Gibbs sampling.

\begin{figure}
\centerline{\includegraphics[width=3.0in]{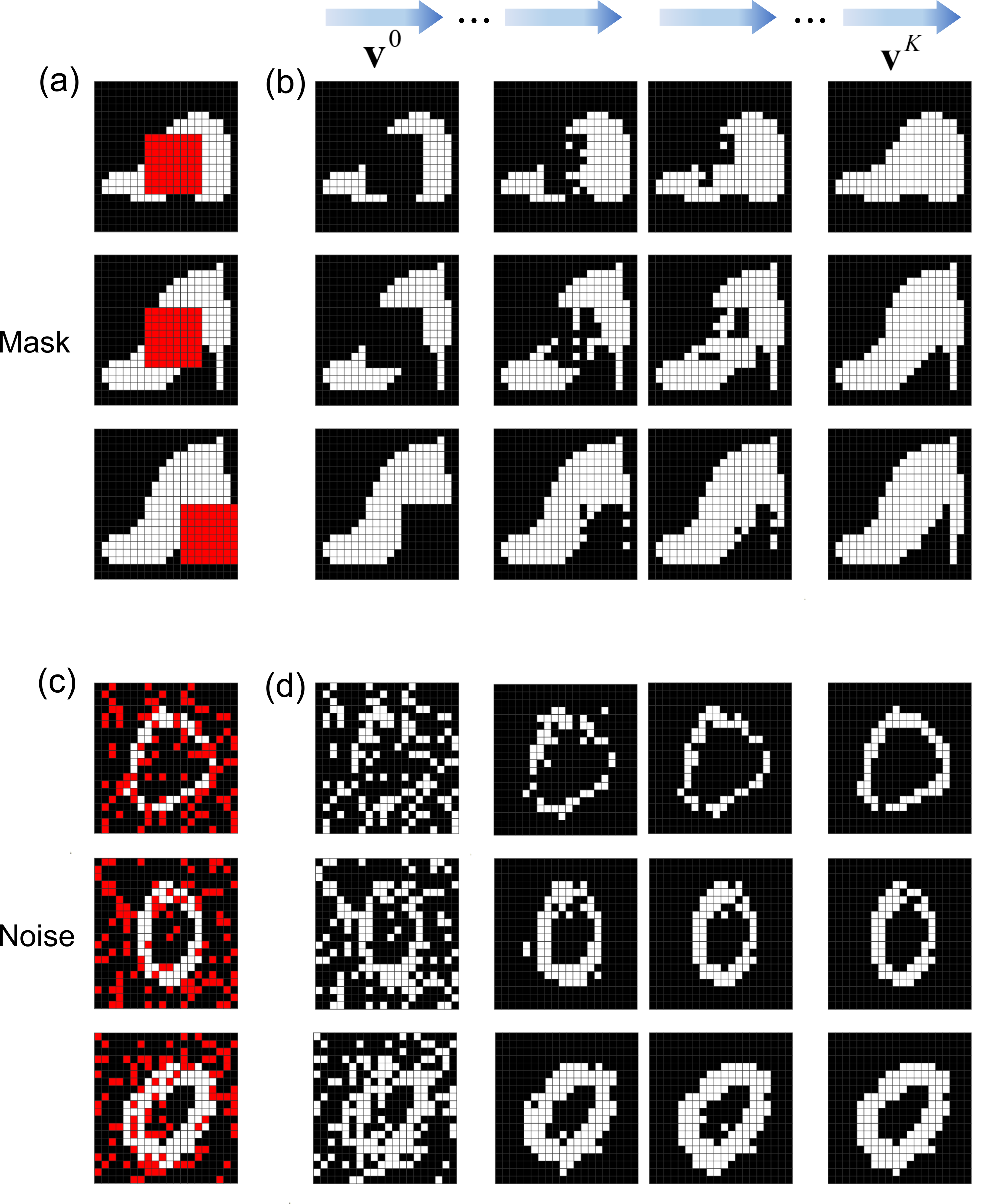}}
\caption{\label{fig:4}\textbf{Image restoration with PRBM.} All the initial images are excluded in the training dataset and distorted. (a) and (b) The processes of image restoration for `Boot' in test dataset are demonstrated, and a portion of the image is occluded (marked by red color). (c) and (d) The processes of image restoration for digit `0' in test dataset are demonstrated, and random noise (marked by red color) is added to digit images.}
\end{figure}

\section{Temporal content generation by PRBM}
To demonstrate time-varying RBM systems, we generate pieces of piano music by using Recurrent Neural Network-Restricted Boltzmann Machine (RNN-RBM) \cite{boulanger2012modeling}, which forms a sequential step through each RBM cell (the green box in Figure 5(a)) along the temporal dimension. Different to the ordinary RBM, the magnetic field coefficients in RNN-RBM are dynamically updated at each time step, whereas the interactions parameters remain invariant. In the experimental implementation of piano music generation with the PRBM, as the time step $t$ increases, we only need to modify the rotation ${\beta _k}$ in Region ${\rm III}$ of the SLM by utilizing the Gauge transform. We use the piano dataset Nottingham \cite{nottingham} as the training dataset. Within the PRBM framework, a total of 88 piano keys ranging from A0 to C8 are mapped onto 88 spins $v_i$ in the visible layer of RBM, where $v_i$ assumes the value of +1 to represent the pressing of the $i$-th piano key and $-1$ otherwise.

Figure 5(a) demonstrates the music generation process. The gray arrows indicate the connections between the elements. The red arrows signify that the spins ${\bf{v}}_t^K$ in each RBM cell correspond to the specific piano keys requiring activation at the current time step. In the experiments, the proposed PRBM executes the music generation composed of 150 time steps, where each time step stands for a quaver. The architecture of each RBM cell within this temporal sequence incorporates 88 spins in the visible layer, corresponding to the full range of piano keys, and 96 spins in the hidden layer to capture the underlying musical patterns and dependencies.

Figure 5(b) presents one of the pieces in training dataset. Following the training process, the rhythmic patterns and stylistic characteristics of the music are encoded within the coefficient parameters. Subsequently, we generate new melodies exhibiting similar rhythmic and stylistic properties. For each time step $t$, ${\bf{v}}_t^0$ is randomly initialized as the spins in visible layer of RBM. The generated music ${\bf{v}}_t^K$ is produced after 20 iterations, as shown in the $t$-th column in the Fig. 5(c). The ${\bf{v}}_t^K$ is then utilized for updating the RNN elements ${{\mathbf{r}}^t}$ in current time step. The generation processes are conducted repeatedly, and after 150 time steps, a piano music piece is generated as shown in Fig.5(c). The generated musical composition demonstrates remarkable similarity to the training examples in terms of rhythmic structure and stylistic attributes, thereby validating the efficacy of the proposed PRBM for temporal content generation.
\begin{figure}
\centerline{\includegraphics[width=3.2in]{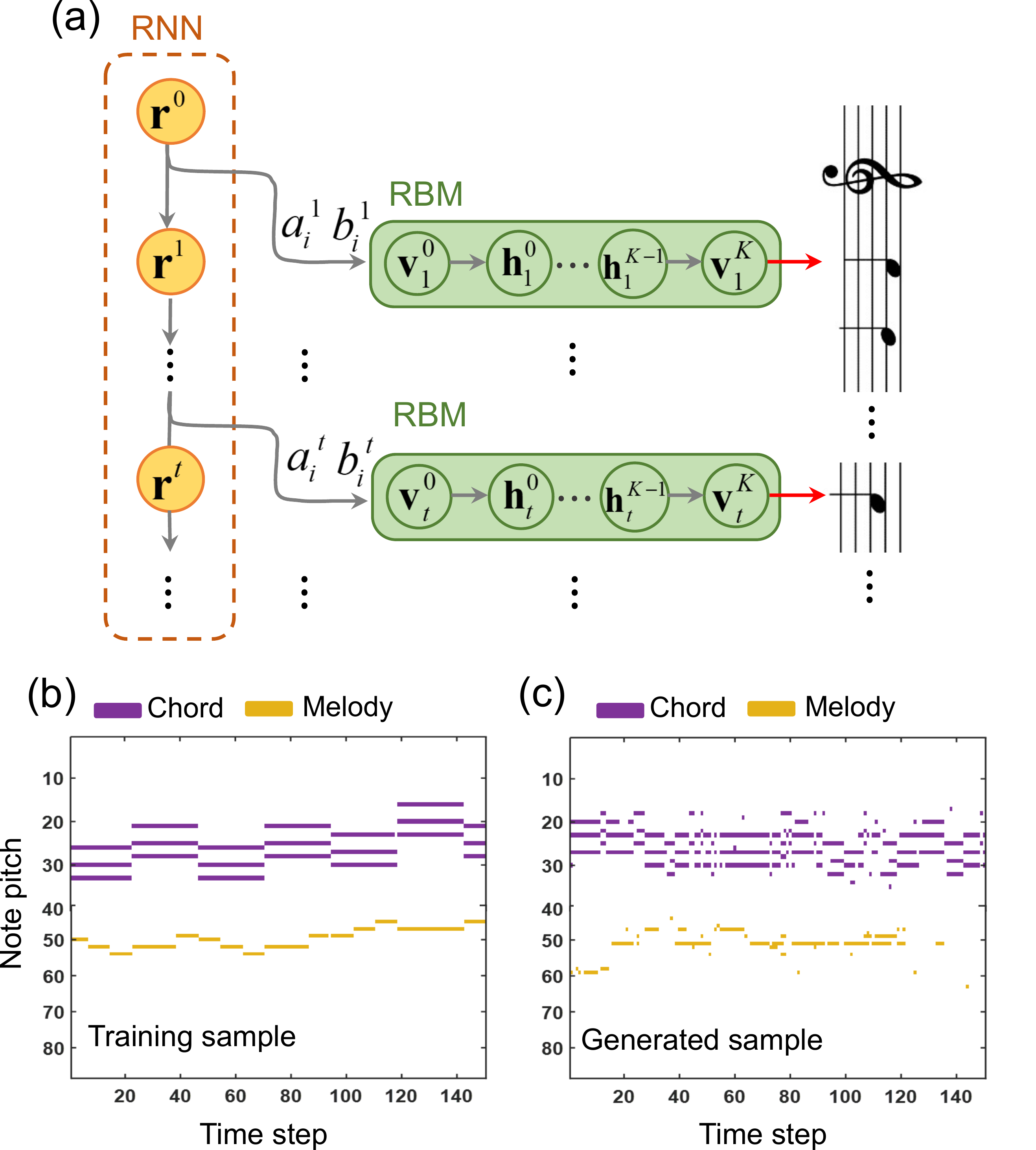}}
\caption{\label{fig:5}\textbf{Temporal content generation with PRBM.} (a) The structure of RNN-RBM. A green box denotes a RBM cell, and the visible layer spin configuration obtained after several iterations is mapped to the piano keys to be pressed at the current time step. (b) A music training sample. (c) The generated music piece.}
\end{figure}

\section{Discussion and Outlook}
In conclusion, we propose a photonic restricted Boltzmann machine with introducing an encoding method to accelerate Gibbs sampling processes. In contrast to SPIM, the proposed encoding method eliminates the decomposition computation of interaction matrix and magnetic field, thereby significantly improving computing efficiency. We verify the PRBM and the photonic computing of Gibbs sampling by observing the phase transition in the 2D lattice lsing model. Furthermore, we experimentally demonstrate the capability of PRBM to generate new content by generating and restoring images. To further illustrate the advantages of the encoding method, we demonstrate that PRBM can implement time-varying spin interactions for temporal content generation, such as music, where non-decomposition computation is particularly crucial for efficient processing.

We note that the proposed system effectively accelerates computation through parallel processing and reduces the computational complexity of Gibbs sampling from $O(N)$ in the traditional electronic computing to $O(1)$ in the photonic computing. Moreover, with the non-Von Neumann photonic computing architecture, the interaction matrix and the magnetic fields are encoded on the SLM, significantly saving the memory cost and breaking the data transportation bottleneck between CPU/GPU and memory in the traditional electronic computation, especially substantially enhancing computation efficiency for large-scale RBM.

We also evaluate the computing capacity and scalability of the proposed system. For each photonic computing of Gibbs sampling, the computational load is represented by Equation (2), which involves $N$ real-number multiplication and $N$ summation, where $N$ is the number of spins. The time per step (TPS) in the present experiment is primarily limited by the modulation speed of the liquid-crystal SLM. With the development of various cutting-edge SLMs capable of gigahertz modulation rate and low power consumption\cite{smolyaninov2019programmable, panuski2022full, li2020lithium}, the minimum TPS could scale down to  ${10^{ - 9}}$s, as PRBM requires only a single pixel photodetector. Moreover, the PRBM can be scaled to larger spin size by extending the wavelength range and increasing the number of SLM pixels. As the pixel scale of the modulator reaches the wavelength scale \cite{li2020lithium}, it is possible to achieve $10^{10}$ pixels in a $100\rm{mm} \times 100\rm{mm}$ area, enabling the realization of models with 10 billion parameters. In this case, the number of spins is $N = 10^5$, the proposed experimental framework can achieve a maximum of 200-tera-floating-point operations per second (TFLOPS).

It is worth noting that using the proposed photonic Gibbs sampling method can potentially improve computing efficiency and reduce resource consumption for training processes. Current generative AI requires tremendous computation resources for training. For example, Transformer-based language models using digital computers require computational operations on the order of $6MBS$, where $M$, $B$ and $S$ are the number of model parameters, the training batch size and the iteration number, respectively \cite{kaplan2020scaling}. In comparison, the total computation operations using digital computers for RBM are the same order of magnitude as that of large language models, approximately $9MBS$, where $M \approx {N^2}/4$, $N$ is the total spin number (Supplementary Materials Section $\rm {V}$). Owing to the accelerated Gibbs sampling enabled by the proposed PRBM, the complexity can be reduced to $3MBS + 3NBS$. Therefore, photonic computing acceleration enables the training efficiency of RBM outperforms the Transformer-based models, which is a significant cue to develop unconventional computational architectures for generative AI.

With the spin size $N$ increasing, the operations completed per second grow, which indicates that the computational capacity and training efficiency of the proposed PRBM could outperform electronic computing. Supplementary Materials Table S1 shows that PRBM reduces the training time by about two orders of magnitude compared to the NVIDIA H100 when training the GPT-3 model with the same number of parameters and training data. Given that the proposed photonic computing scheme demonstrates strong compatibility with diverse data types, including both sequential and non-sequential data, we believe that PRBM could achieve high practical efficiency for learning complex probability distributions and generating sophisticated contexts, including potential applications in language models.

\section*{Methods}
\textbf{Restricted Boltzmann machine} At temperature $T$, the probability of a spin configuration follows the Boltzmann-Gibbs distribution $P({\bf{v,h}}) \propto \exp [ - H({\bf{v,h}})/T]$, where $H$ is the energy of that state (here we take the convention of the Boltzmann constant ${k_B} = 1$). As shown in Figure 1(a), the training set is binarized as the spins in the visible layer, and the training objective of RBMs is to maximize the log-likelihood function $\ell ({\bm{\theta }}){\rm{ = }}{\left\langle {\ln P({\bf{v,h}})} \right\rangle _{data}}$ with the parameters ${\bm{\theta }} = \left\{ {{W_{ij}},{a_i},{b_i}} \right\}$, at temperature $T=1$. Gibbs sampling is utilized to compute the partial derivatives of $\ell ({\bm {\theta }})$ with respect to each parameter, which enables the update of parameters. The probability of the  $k$-th spin being $+1$ in the hidden layer is $P({h_k} = 1|{\bf{v}}) = P({h_k} = 1,{\bf{v}})/P({\bf{v}}) = 1/[1 + \exp ( - 2({b_k} + \sum\nolimits_i {{v_i}{W_{ik}}} ))]$ (Supplementary Materials Section $\rm {VI}$), while the probability of the $k$-th spin in the visible layer being $+1$ is calculated as $P({v_k} = 1|{\bf{h}}) = 1/[1 + \exp ( - 2({a_k} + \sum\nolimits_i {{W_{ki}}{h_i}} ))]$. In the generation process, Gibbs sampling is employed to generate new content of the same type based on the trained parameters.

\textbf{General sparse interaction matrix} The proposed encoding method is also applicable to Ising models with complex sparse interactions. During Gibbs sampling, the spin ${\sigma _i}$ is encoded in Region ${\rm I}$ of SLM, while the spins connected to ${\sigma _i}$ are encoded in Region ${\rm II}$. Therefore, this photonic computing system can be extended to simulate general Ising models with sparse interactions. Especially for combinatorial optimization problems, such as Graph Partitioning Problem with arbitrary sparsity, the system is expected to exhibit unique performance without requiring matrix decomposition.

\section*{Acknowledgement}
The authors acknowledge funding through the National Natural Science Foundation of China (124B2085, 12474393 and 12174340), and the National Key Research and Development Program of China (2022YFA1405200).

%\bibliography{citation}

%=======================================

\newpage
\clearpage

\setcounter{section}{0}

\newcommand{\hbAppendixPrefix}{S}
\renewcommand{\thefigure}{\hbAppendixPrefix\arabic{figure}}
\setcounter{figure}{0}

\renewcommand{\thetable}{\hbAppendixPrefix\arabic{table}}
\setcounter{table}{0}
\renewcommand{\theequation}{\hbAppendixPrefix\arabic{equation}}
\setcounter{equation}{0}

\onecolumngrid

\begin{center}
\large{\textbf{
Supplementary Material: Photonic restricted Boltzmann machine for content generation tasks }}
\end{center}

\section{Experimental set-up for photonic restricted Boltzmann machine}

In Figure S1, a supercontinuum laser source (Anyang SC-5) produces a collimated Gaussian beam. The beam waist radius undergoes tenfold expansion through a lens system comprising L1 (50 mm in focal length) and L2 (500 mm in focal length). The expanded beam subsequently illuminates a reflective diffraction grating (inscribed line density of 600/mm), and the cylindrical lens (150 mm in focal length) focus the light with different wavelengths onto an SLM along the $x$-axis, while the $y$-axis pixels are illuminated coherently by the same wavelength. Polarizer P ensures linear polarization alignment with the SLM's long display axis. L3 (100 mm in focal length) and L4 (200 mm in focal length) magnify twofold at the SLM plane (Holoeye PLUTONIR-011). Lens FL (200 mm in focal length) performs a Fourier transformation of the optical field, resulting in an incoherent intensity summation of different wavelengths and coherent interference for each wavelength at the back focus plane. An sCMOS camera (Andor Zyla 5.5) positioned at this focal plane records the intensity distribution, following the measurement and feedback scheme. To account for finite SLM pixel dimensions and sensor resolution limitations, intensity measurements are obtained through spatial integration around central regions rather than single-point detection. In the experiment, the operational wavelengths are selected from 588 $\rm nm$ to 611 $ \rm nm$. Each individual wavelength projects a beam covering approximately two pixels along the $x$-axis of the SLM, approximately 13.5 $\rm\mu m$. Each pixel of SLM has a physical width of 8 $\rm\mu m$. Based on the grating equation, the wavelength interval for two neighboring wavelengths is estimated about $\Delta \lambda = 0.053 \rm{nm}$.
\begin{figure}[h]
\centerline{\includegraphics[width=6in]{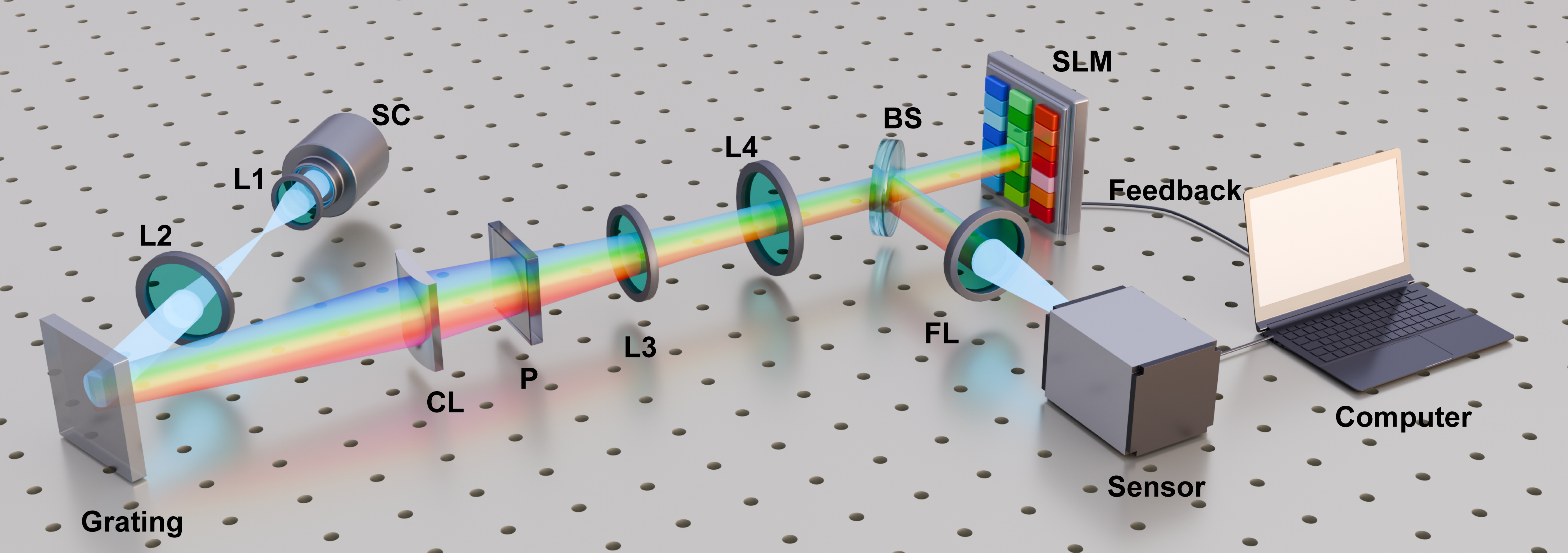}}
\caption{\label{fig:S1}Schematic of experimental set-up for photonic restricted Boltzmann machine. SC, supercontinuum laser; L1, L2, L3, L4, and FL, lens; CL, cylindrical lens; P, polarizer; BS, beam splitter; Sensor, sCMOS camera.}
\end{figure}
\section{Photonic computing of Gibbs sampling}
 We take the example of the generating process as illustrated in Figure 1(c) in the main text. We first assume that the light intensity for each wavelength on the SLM is uniform and equal to $A_0$. When sampling from the visible layer to the hidden layer, for each wavelength, the phase modulation $\phi _{j,k}^{m,n}$ on the SLM is defined as Eq. (1) in the main text. With proposed Gauge transform, each spin encoded by ${N_x} \times {N_y}$ pixels. Therefore, the electric field modulated after SLM is expressed as
\begin{equation}
{E_k}(x,y) = {A_0}[{G_ + } \cdot H + {G_ - } \cdot (1 - H)] \cdot {\rm{Rec}}{{\rm{t}}_{{N_x}W}}(x)\label{eq:S1}
\end{equation}
where ${G_+ }$ and ${G_- }$ are written as
\begin{equation}
\begin{aligned}
{G_ \pm }&= \sum\limits_{j = 1}^{{N_v} + 2} {{e^{ \pm i{\phi _{j,k}}}}{\rm{Rec}}{{\rm{t}}_{{N_y}W}}\left( {y - {y_j}} \right)} \\
 &= i{h_k} \cdot {\rm{Rec}}{{\rm{t}}_{{N_y}W}}\left( {y - {y_1}} \right){\rm{ + }}i\sum\limits_{j = 1}^{{N_v}} {{e^{ \pm i{\alpha _{jk}}}}{v_j} \cdot {\rm{Rec}}{{\rm{t}}_{{N_y}W}}\left( {y - {y_{j + 1}}} \right)}  + i{e^{ \pm i{\beta _k}}}s \cdot {\rm{Rec}}{{\rm{t}}_{{N_y}W}}\left( {y - {y_{{N_v} + 2}}} \right)
\end{aligned}
\end{equation}
In this context, $(x,y)$ represent the spatial coordinate on the SLM, and $y_j=j{{N}_{y}}W$ represents the position of the $j$-th spin in $y$ direction on SLM. The rectangular function is defined as
\begin{equation}
{{\mathop{\rm Rect}\nolimits} _{{N_y}W}}(y) ={\rm{Rect}}(\frac{y}{N_yW})= \left\{ {\begin{array}{*{20}{l}} {1}, &{|y| \le {N_y}W/2}\\ 0, &{|y| > {N_y}W/2} \end{array}} \right.
\end{equation}
where $W$ represents the pixel width of the SLM. $H$ is the checkerboard function, so $1 - H$ is the complementary function. The checkerboard function is defined as
\begin{equation}
\begin{array}{c}
H = \sum\limits_{m,n =  - \infty }^\infty  {[\delta (x - 2mW,y - 2nW) + \delta (x - (2m + 1)W,y - (2n + 1)W)]}  \otimes {\rm{Rec}}{{\rm{t}}_W}(x,y)
\end{array}
\end{equation}
where $m$ and $n$ are two integers, and the mathematical symbol $\otimes $ denotes the convolution operation. The optical field at the focal plane after passing through the lens is expressed as the result of a two-dimensional Fourier transform, written as follows
\begin{equation}
{\tilde E_k}({k_x},{k_y}) = \frac{1}{{i\lambda f}}\frac{{{A_0}}}{4}({P_ + } \otimes {Q_1} + {P_ - } \otimes {Q_2})
\end{equation}
where
\begin{equation}
{P_ \pm } = \sum\limits_{j = 1}^{{N_v} + 2} {{e^{ \pm i{\phi _{j,k}}}}}  \cdot {\rm{sin}}{{\rm{c}}_{{N_y}W}}\left( {{k_y}} \right){e^{i{k_y}{y_j}}} \cdot ({N_y}W) \cdot {\rm{sin}}{{\rm{c}}_{{N_x}W}}({k_x}) \cdot ({N_x}{\rm{W}})
\end{equation}
\begin{equation}
{Q_1} = \sum\limits_{m,n =  - \infty }^\infty  {\left( {1 + {{( - 1)}^{m + n}}} \right)} {\mkern 1mu} \delta \left( {{k_x} - m\frac{\pi }{W},{k_y} - n\frac{\pi }{W}} \right) \cdot {\rm{sin}}{{\rm{c}}_W}({k_x},{k_y}) \notag
\end{equation}
\begin{equation}
{Q_2} = \sum\limits_{m,n =  - \infty }^\infty  {\left( {{{( - 1)}^m} + {{( - 1)}^n}} \right)} {\mkern 1mu} \delta \left( {{k_x} - m\frac{\pi }{W},{k_y} - n\frac{\pi }{W}} \right) \cdot {\rm{sin}}{{\rm{c}}_W}\left( {{k_x},{k_y}} \right) \notag
\end{equation}
\begin{equation}
{\rm{sin}}{{\rm{c}}_W}({k_x},{k_y}) = {\rm{sinc}}(\frac{{{k_x}W}}{{2\pi }}) \cdot {\rm{sinc}}(\frac{{{k_y}W}}{{2\pi }}) \notag
\end{equation}
\begin{equation}
{\rm{sinc}}(k) = \frac{{{\rm{sin}}(k\pi )}}{{k\pi }} \notag
\end{equation}
Due to the checkerboard modulation, the optical beam is diffracted into multiple orders at $(m\frac{\pi }{W}, n\frac{\pi }{W})$ in the angular spectral space. Furthermore, we convert the angular spectrum coordinates $({u_x},{u_y})$ of the detection area with ${u_x} = {k_x}\frac{{f\lambda }}{{2\pi }}$ and ${u_y} ={k_y}\frac{{f\lambda }}{{2\pi }}$. Due to the minimal overlap of the light fields between different orders, it can be neglected. The zeroth-order diffraction field, where both $m$ and $n$ are equal to zero, can be expressed as
\begin{equation}
\begin{aligned}
{\tilde E_k}({u_x},{u_y}) \buildrel\textstyle.\over=   \frac{{{A_0}C}}{{\lambda f}}{\rm{(}}{h_k}{e^{i\frac{{2\pi }}{{\lambda f}}{u_y}{y_1}}}{\rm{ + }}\sum\limits_{j = 1}^{{N_v}} {{W_{jk}}{v_j}{\rm{/}}L}  \cdot {e^{i\frac{{2\pi }}{{\lambda f}}{u_y}{y_{j+1}}}} + {b_k}{\rm{/}}L \cdot {e^{i\frac{{2\pi }}{{\lambda f}}{u_y}{y_{{N_v} + 2}}}}{\rm{)sin}}{{\rm{c}}_W}(0,0){\rm{sinc}}\left( {\frac{{{u_x}{N_x}W}}{{\lambda f}}} \right){\rm{sinc}}\left( {\frac{{{u_y}{N_y}W}}{{\lambda f}}} \right) \notag
\end{aligned}
\end{equation}
where $C={N}_{x}W\cdot {{N}_{y}}W$. Therefore, the detected intensity at ${u_x} = 0$ and ${u_y} = 0$ is
\begin{equation}
\begin{aligned}
{I_k}(0,0) = \frac{{{{({A_0}C)}^2}}}{{{{(\lambda f)}^2}}}{\rm{(}}{h_k}{\rm{ + }}{b_k}{\rm{/}}L{\rm{ + }}\sum\limits_{j = 1}^{{N_v}} {{W_{jk}}{v_j}{\rm{/}}L{{\rm{)}}^2}}
\end{aligned}
\end{equation}
Owing to the incoherence among different wavelengths, the overall detection intensity at the zeroth order is given by
\begin{equation}
I = \sum\limits_k^{{N_h}} {{I_k}} (0,0) = \frac{{{{({A_0}C)}^2}}}{{{{(\lambda f)}^2}}}\sum\limits_{k = 1}^{{N_h}} {{\rm{(}}{h_k}{\rm{ + }}{b_k}{\rm{/}}L{\rm{ + }}\sum\limits_{j = 1}^{{N_v}} {{W_{jk}}{v_j}{\rm{/}}L{{\rm{)}}^2}} }
\end{equation}
where $N_h$ is the number of the spins in the hidden layer. We define all coefficients and spins to be equal to 1, with detected intensity ${I_{norm}} = \frac{{{{({A_0}C)}^2}}}{{{{(\lambda f)}^2}}}{N_h}{({N_v}/L + 1/L + 1)^2}$. By normalizing the detected intensity to $I_{norm}$, the normalized intensity $\tilde I$ is written as
\begin{equation}
\begin{aligned}
\tilde I = {N_h}{({N_v}/L + 1/L + 1)^2}\frac{I}{{{I_{norm}}}} = \sum\limits_{k = 1}^{{N_h}} {{\rm{(}}{h_k}{\rm{ + }}{b_k}{\rm{/}}L{\rm{ + }}\sum\limits_{j = 1}^{{N_v}} {{W_{jk}}{v_j}{\rm{/}}L{{\rm{)}}^2}} }
\end{aligned}
\end{equation}

Then we take the example of the generating process, we perform photonic computing of Gibbs sampling in two steps. For example, when sampling from the visible layer to obtain the hidden layer, in the first step, we set all hidden layer spins ${h_i} = 1$ and encode them with a random initial state onto the SLM following Equation (1) in the main text. According to Eq.(S8) and Eq.(S9), the normalized intensity ${\tilde I_0}$ at the center position of the back focus plane is obtained as ${\tilde I_0} = \sum\limits_{k = 1}^{{N_h}} {{\rm{(1 + }}{b_k}{\rm{/}}L{\rm{ + }}\sum\limits_{j = 1}^{{N_v}} {{W_{jk}}{v_j}{\rm{/}}L{{\rm{)}}^2}} }$. Thus, the Hamiltonian is computed as
\begin{equation}
{H_0} =  - L{\tilde I_0} =  - 2\sum\limits_{i = 1}^{{N_h}} {\sum\limits_{j = 1}^{{N_v}} {{v_j}{W_{ji}} - } } 2\sum\limits_{i = 1}^{{N_h}} {{b_i} - } {N_h}L - \sum\limits_{i = 1}^{{N_h}} {(\sum\limits_{j = 1}^{{N_v}} {{v_j}{W_{ji}}{)^2}} } /L - \sum\limits_{i = 1}^{{N_h}} {{{({b_i})}^2}/L - } 2\sum\limits_{i = 1}^{{N_h}} {\sum\limits_{j = 1}^{{N_v}} {{v_j}{W_{ji}}{b_i}} } /L
\end{equation}
When calculating the Gibbs sampling probability for the hidden layer spins, the visible layer spins and the magnetic field remain unchanged. Thus, the above expression simplifies to
\begin{equation}
{H_0} =  - 2\sum\limits_{i = 1}^{{N_h}} {\sum\limits_{j = 1}^{{N_v}} {{v_j}{W_{ji}} - } } 2\sum\limits_{i = 1}^{{N_h}} {{b_i} - } Q
\end{equation}
where $Q$ is a constant. In the second step, each spin in the hidden layer is flipped independently. For the $k$-th spin, we set ${h_k} =  - 1$ and compute the corresponding Hamiltonian as
\begin{equation}
{H_k} =  - 2\sum\limits_{i \ne k}^{{N_h}} {\sum\limits_{j = 1}^{{N_v}} {{v_j}{W_{ji}} - } } 2\sum\limits_{i \ne k}^{{N_h}} {{b_i} + 2\sum\limits_{j = 1}^{{N_v}} {{v_j}{W_{jk}}}  + 2{b_k} - Q}
\end{equation}
Thus, the Hamiltonian difference is $\Delta H_k = ({H_0} - {H_k})/2 =  - 2\sum\limits_{j = 1}^{{N_v}} {{v_j}{W_{jk}}}  - 2{b_k}$, and the Gibbs sampling probability is $P({h_k} = 1|{\bf{v}}) = 1/(1 + {{\mathop{\rm e}\nolimits} ^{\Delta H_k}})$. The photonic computing system we propose reduces the computational complexity of Gibbs sampling from $O(N)$ to $O(1)$, substantially enhancing computation efficiency for large-scale RBM.

\section{Ground-state search for two-dimensional Ising model}
Figure S2 illustrates the variation of magnetization $M$ (orange line) and Hamiltonian $H$ (blue line) during the cooling process from high to low temperature. The black line denotes the temperature variation during the iterative process, which undergoes a decay from an initial temperature $T=6.1J$ to a final temperature $T=0.3J$. The Hamiltonian gradually decreases with decreasing temperature, eventually reaching its minimum value. During the cooling process, the magnetization shows that the spin configuration transitions from a paramagnetic phase to a ferromagnetic phase.

\section{Iterative generating with photonic computing of Gibbs sampling }
In the image generation experiments, we shrink the original dataset to $14\times14$ pixels and binarize them as the visible layer spins. We set the number of hidden layer spins as 196. To train the model, we use the contrastive divergence (CD-1) algorithm and train the model in the absence of magnetic field. We set the learning rate for the interaction parameters to 0.003, and to avoid overfitting, the weight decay parameter is set to 0.0002. The initial values of the interaction parameters ${W_{ij}}$ follow a Gaussian distribution with a mean of 0 and a variance of 0.01. We train on 1000 images, dividing them into 50 batches for training. The total number of training epochs is 2000. During the training process, we introduce the Momentum method to address issues such as slow convergence, oscillations, and getting trapped in local optima in the gradient descent algorithm. Figure S3(a) shows the evolution of the log-likelihood function in the training process for item `Boot'. The log-likelihood function gradually increases and approaches 0 as the number of training iterations increases, demonstrating that our training is effective. In the generation of the image, an initial state is randomly generated, and Gibbs sampling is performed for each spin of the hidden and visible layers in turn, and so on for 15 iterations, with the final visible layer spins representing the generated image.

\begin{figure}[t]
\centerline{\includegraphics[width=3in]{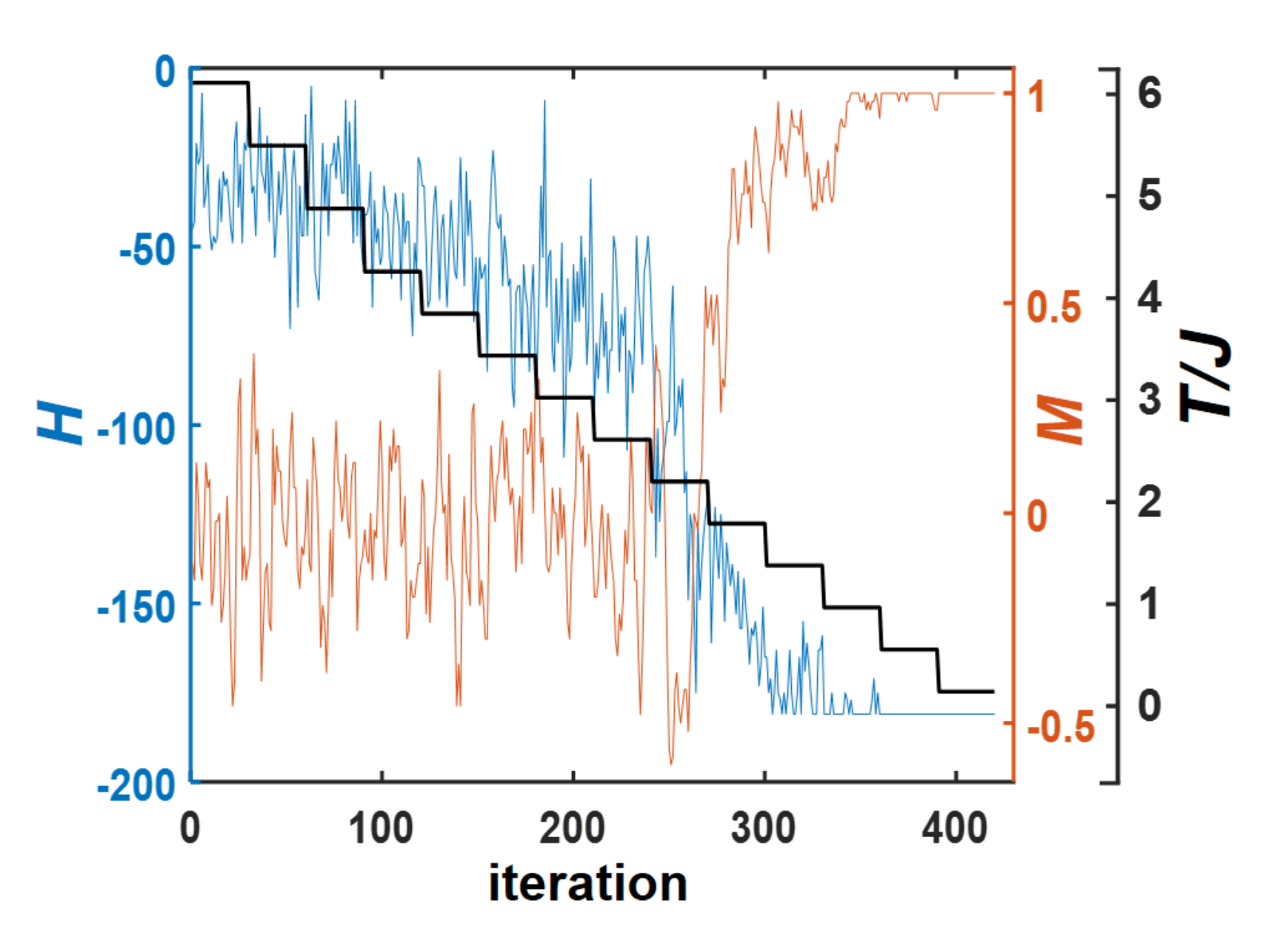}}
\caption{\label{fig:S2}The variation of the magnetization $M$ (orange line) and Hamiltonian $H$ (blue line) of the two-dimensional Ising model during the gradual cooling process from high temperature to low temperature.}
\end{figure}

In the experiments of image restoration, we shrink the original dataset to $20\times20$ pixels and binarize them as the visible layer spins. The number of hidden layer spins is same to that of visible layer spins. In the process of learning the parameters, we employ the same learning rate and initial values for the parameters as used in the generation experiments. We train on 1500 images, dividing them into 50 batches for training. The total number of training epochs is 2000. Figure S3(b) shows the evolution of the log-likelihood function in the training process for item `Boot'. To test over-fitting, we use different damaged validation images as the initial states and as the spins of visible layer. The Gibbs sampling is performed for each spin of the hidden and visible layers in turn, and so on for 15 iterations, with the final visible layer spins representing the restored images.

For temporal content generation in the PRBM, we use the piano dataset (Nottingham) as the training dataset. The RNN-RBM model is trained with CD-15 algorithm to maximize the log-likelihood function, and we introduce the Adam optimization algorithm during the training process. All parameters are initialized according to a Gaussian distribution with a mean of 0 and a variance of 0.01, and the learning rate is set to 0.001. We train on 1000 piano songs, dividing them into 10 batches for training. The total number of training epochs is 200, so the training iteration is 2000. Figure S3(c) illustrates the evolution of the log-likelihood function.

\begin{figure}[h]
\centerline{\includegraphics[width=6in]{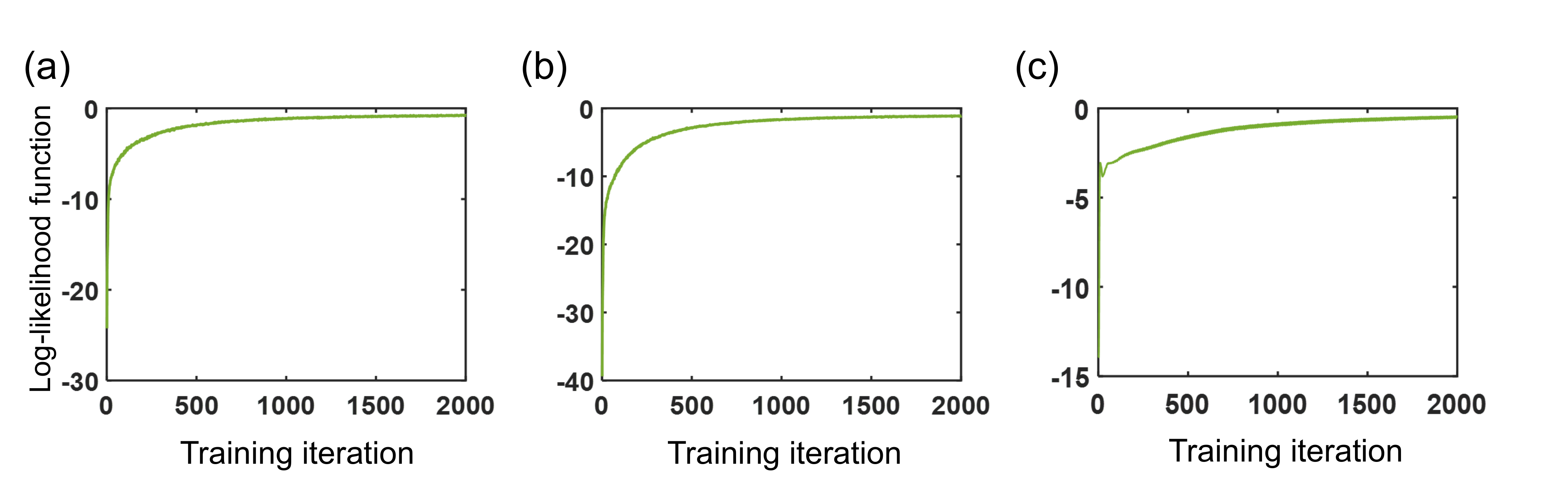}}
\caption{\label{fig:S3}The variation of the log-likelihood function with respect to the number of training iterations. The log-likelihood function gradually increases and approaches 0, demonstrating that our training is effective. (a) is the log-likelihood function for item `Boot' in the image generation experiments. (b) is the log-likelihood function for item `Boot' in the image restoration experiments. (c) is the log-likelihood function in the music generation experiments.}
\end{figure}

\section{Derivation of the digital computational complexity for training RBM}
Training an RBM typically uses the CD-$k$ (Contrastive Divergence) algorithm, where in most cases $k = 1$. Here, we also use $k = 1$ to calculate the computational complexity. Assuming both the hidden layer and visible layer have $N/2$ spins, and the number of training parameters including interactions and magnetic field is $M$, where $M = {N^2}/4 + N$. Since the number of magnetic field is negligible compared to the number of interactions, we define $M \approx {N^2}/4$. For given visible layer, calculating the probability of each spin in the hidden layer using Gibbs sampling requires $N/2$ real-number multiplication, $N/2$ summation and one computation of the sigmoid function. Since the computational cost of the sigmoid function is very small compared to the multiplication and summation, it can be neglected. Therefore, the computational complexity for calculating all the spins in the hidden layer is $N \times N/2$. Similarly, the computational complexity for reconstructing the visible layer from the hidden layer is $N \times N/2$. Based on CD-1 algorithm, an additional computation is required to sample the hidden layer from the reconstructed visible layer, which also has a computational complexity of $N \times N/2$. Assuming that the batch size for each parameter update is $B$, the total digital computational complexity required for Gibbs sampling is the summation of the aforementioned calculations, which yields a final complexity of $3{N^2}BS/2$ with total training iteration $S$.

When calculating the gradient of the interactions for batch size $B$, $2MB$ multiplication, and $MB$ summation operations are required in each iteration. The computation of the magnetic field gradient requires only $2NB$ summation, which is negligible compared to the computational cost of the interactions. Therefore, the total computation complexity required to train the RBM is approximately $C \approx 9MBS$, and Gibbs sampling constitutes a significant portion of the total computational complexity. With the proposed PRBM, the computational complexity of Gibbs sampling is reduced to $3NBS$. Therefore, the total computational complexity for training the RBM is $3MBS + 3NBS$.

\section{Derivation of Gibbs sampling}

In general, Gibbs sampling involves the visible and hidden layer spins taking values of 0 or 1. However, given that the spin in Ising model is -1 or 1, the original formula for calculating the Gibbs sampling probability cannot be applied to the system proposed in this article. Therefore, we present the derivations of Gibbs sampling in the proposed system.

Given the Hamiltonian $H$, the joint probability distribution of the model is represented as
\begin{equation}
p({\mathbf{v,h}}) = \frac{1}{Z}\exp [ - H({\mathbf{v,h}})/T],\;\;\;Z = \sum\limits_{{\mathbf{v,h}}} {\exp [ - H({\mathbf{v,h}})/T]}.
\end{equation}
Furthermore, we set the temperature $T$ in the experiment at 1. In the process of Gibbs sampling, the conditional probability of each spin is given by
\begin{equation}
p({v_i}|{{\mathbf{v}}_{\backslash i}},{\mathbf{h}}) = p({v_i}|{\mathbf{h}}),\quad p({h_j}|{\mathbf{v}},{{\mathbf{h}}_{\backslash j}}) = p({h_j}|{\mathbf{v}})
\end{equation}
It is known from the conditional probability formula that $p({h_j} = 1|{\mathbf{v}}) = \frac{{p({h_j} = 1,{\mathbf{v}})}}{{p({\mathbf{v}})}}$. The marginal probability can be expressed as
\begin{equation}
\begin{array} {ll}
  p({\mathbf{v}})&= \sum\limits_{\mathbf{h}} {p({\mathbf{v,h}})}  = \frac{1}{Z}\sum\limits_{\mathbf{h}} {\exp [ - H({\mathbf{v,h}})} ] \\
   &= \frac{1}{Z}\sum\limits_{\mathbf{h}} {\exp [\sum\limits_{ik} {{v_i}{W_{ik}}{h_k} + \sum\limits_i {{a_i}{v_i}}  + } \sum\limits_i {{b_i}{h_i}} } ] \\
   &= \frac{{\exp ({a^T})}}{Z}\sum\limits_{\mathbf{h}} {\exp [\sum\limits_k {{h_k}(\sum\limits_i {{v_i}{W_{ik}}}  + } {b_k}} )] \\
  & = \frac{{\exp ({a^T})}}{Z}\sum\limits_{\mathbf{h}} {\prod\limits_k {\exp [{h_k}(\sum\limits_i {{v_i}{W_{ik}}}  + {b_k})]} }  \\
   &= \frac{{\exp ({a^T})}}{Z}\sum\limits_{{h_1}} {\sum\limits_{{h_2}} { \cdots \sum\limits_{{h_n}} {\prod\limits_k {\exp [{h_k}(\sum\limits_i {{v_i}{W_{ik}}}  + {b_k})]} } } }  \\
   &= \frac{{\exp ({a^T})}}{Z}\prod\limits_k {\sum\limits_{{h_k}} {\exp [{h_k}(\sum\limits_i {{v_i}{W_{ik}}}  + {b_k})]} }  \\
\end{array}
\end{equation}
So
\begin{equation}
p({h_j} = 1,{\mathbf{v}}) = \frac{{\exp ({a^T})}}{Z}\prod\limits_{k \ne j} {\{ \sum\limits_{{h_k}} {\exp [{h_k}(\sum\limits_i {{v_i}{W_{ik}}}  + {b_k})]} } \} \exp (\sum\limits_i {{v_i}{W_{ij}}}  + {b_j})
\end{equation}
Therefore
\begin{equation}
p({h_j} = 1|{\mathbf{v}}) = \frac{{\exp (\sum\limits_i {{v_i}{W_{ij}}}  + {b_j})}}{{\exp (\sum\limits_i {{v_i}{W_{ij}}}  + {b_j}) + \exp ( - \sum\limits_i {{v_i}{W_{ij}}}  - {b_j})}} = \frac{1}{{1 + \exp ( - 2\sum\limits_i {{v_i}{W_{ij}}}  - 2{b_j})}}
\end{equation}
Similarly
\begin{equation}
p({v_i} = 1|{\mathbf{h}}) = \frac{1}{{1 + \exp ( - 2\sum\limits_j {{h_j}{W_{ij}}}  - 2{a_i})}}
\end{equation}

%\newpage

\section{Experimental results on image generation and restoration}
We have trained the RBM on all digit classes from the MNIST dataset, as well as the `Boot', `Pants', and `T-shirt' categories from the Fashion-MNIST dataset. The experimental generation results for digit `0', `Boot', and `Pants', obtained through our proposed PRBM and encoding method, are presented in the main text. The following figures demonstrate the experimental generation processes for the remaining image categories. In Figures S4-S13, the first column displays randomly initialized spin configuration, the subsequent columns (from left to right) illustrate the iterative stages of the image generation process, and the final column presents the generated image. Figure S14 demonstrates the image restoration process with three different images, which are sourced from the test set. The first column displays the corrupted image with a masked region, which serves as input for the restoration experiments. Similarly, Figure S15 demonstrates the restoration process for three images corrupted by random noise.
\vspace{0pt}
\begin{figure}[h]
\centerline{\includegraphics[width=3.9in]{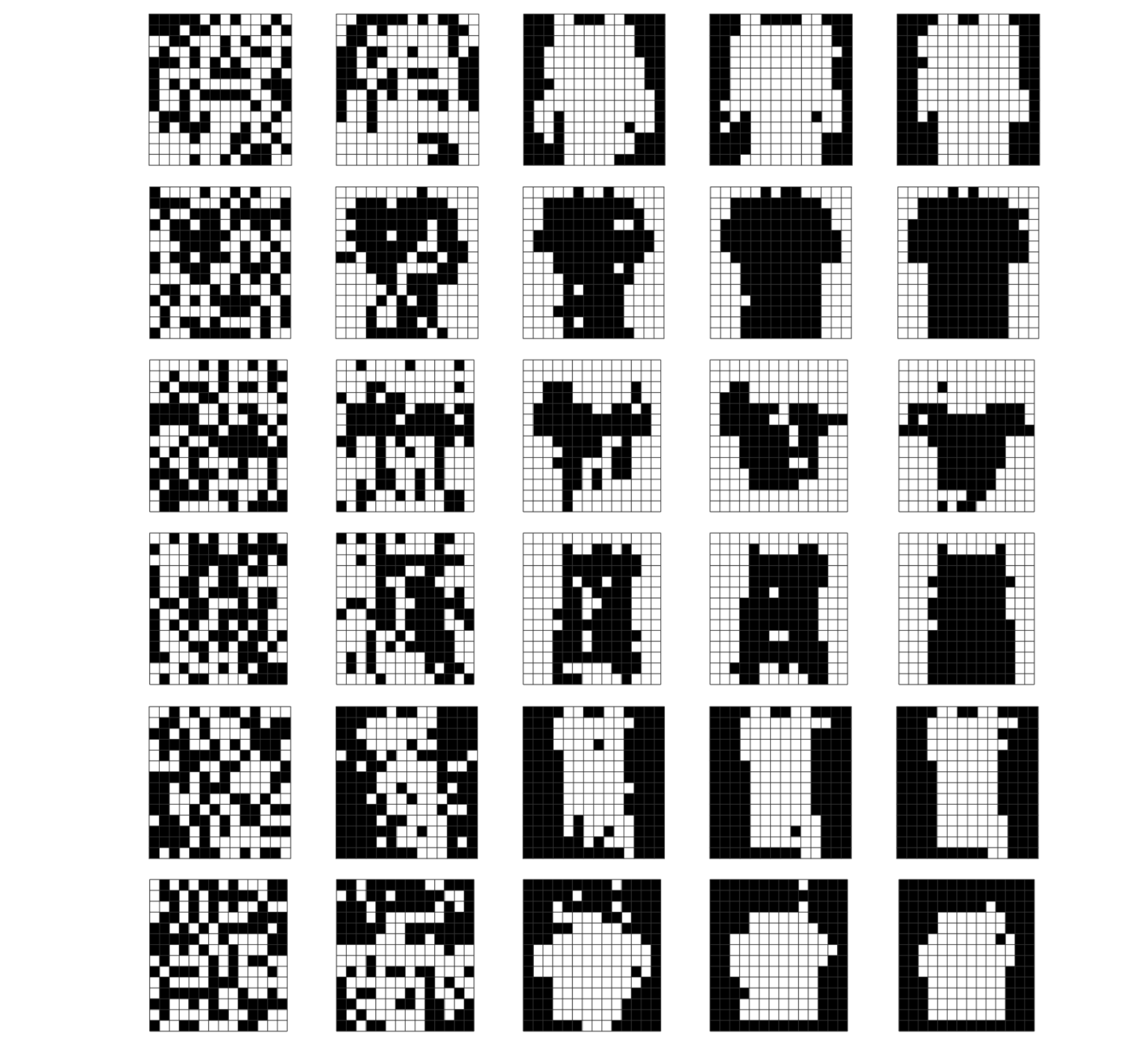}}
\caption{\label{fig:S4}The processes of generating six independent images for fashion `T-shirt'.}
\end{figure}
\begin{figure}[h]
\centerline{\includegraphics[width=3.9in]{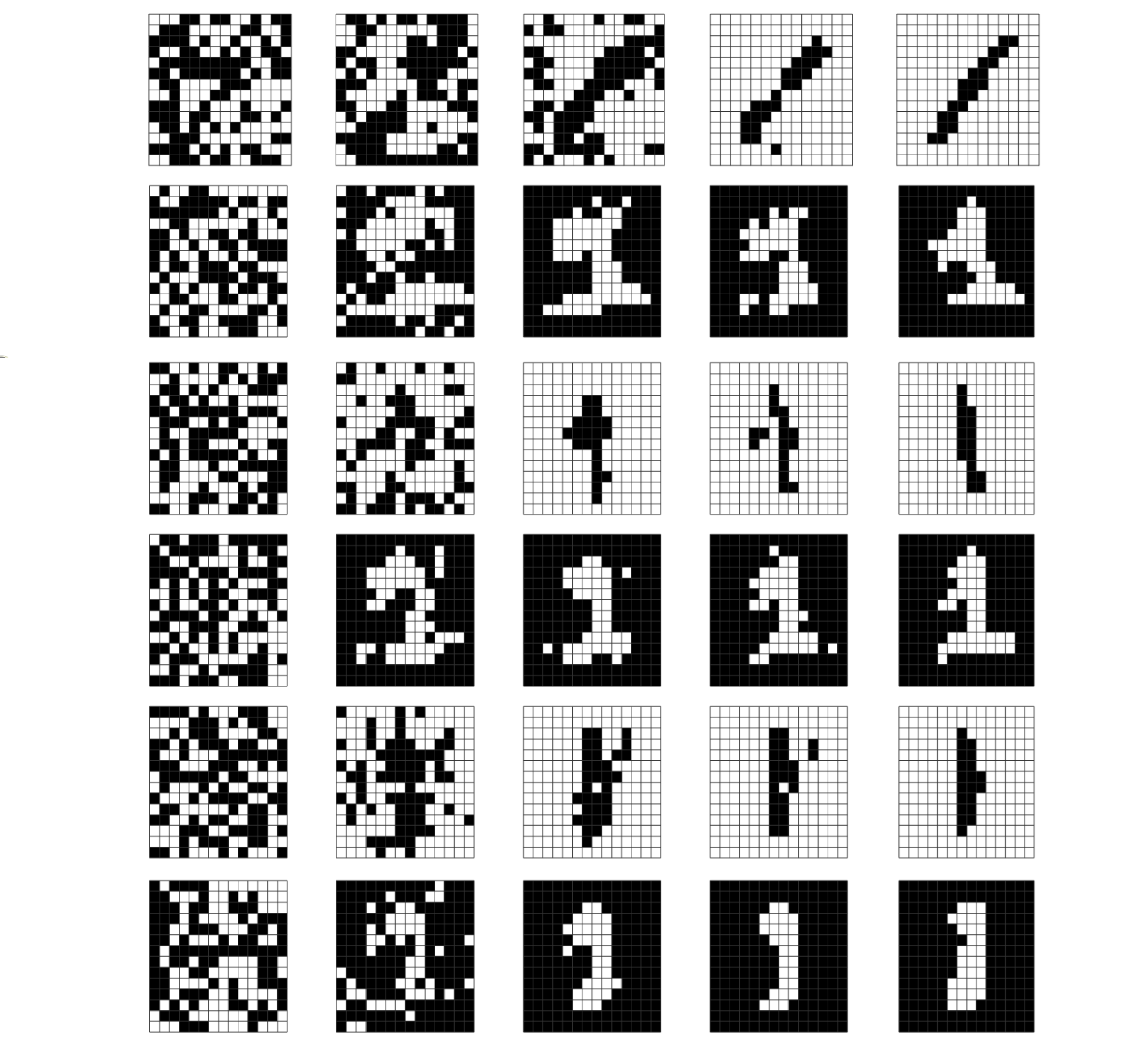}}
\caption{\label{fig:S5}The processes of generating six independent images for the digit `1'.}
\end{figure}
\begin{figure}[h]
\centerline{\includegraphics[width=3.9in]{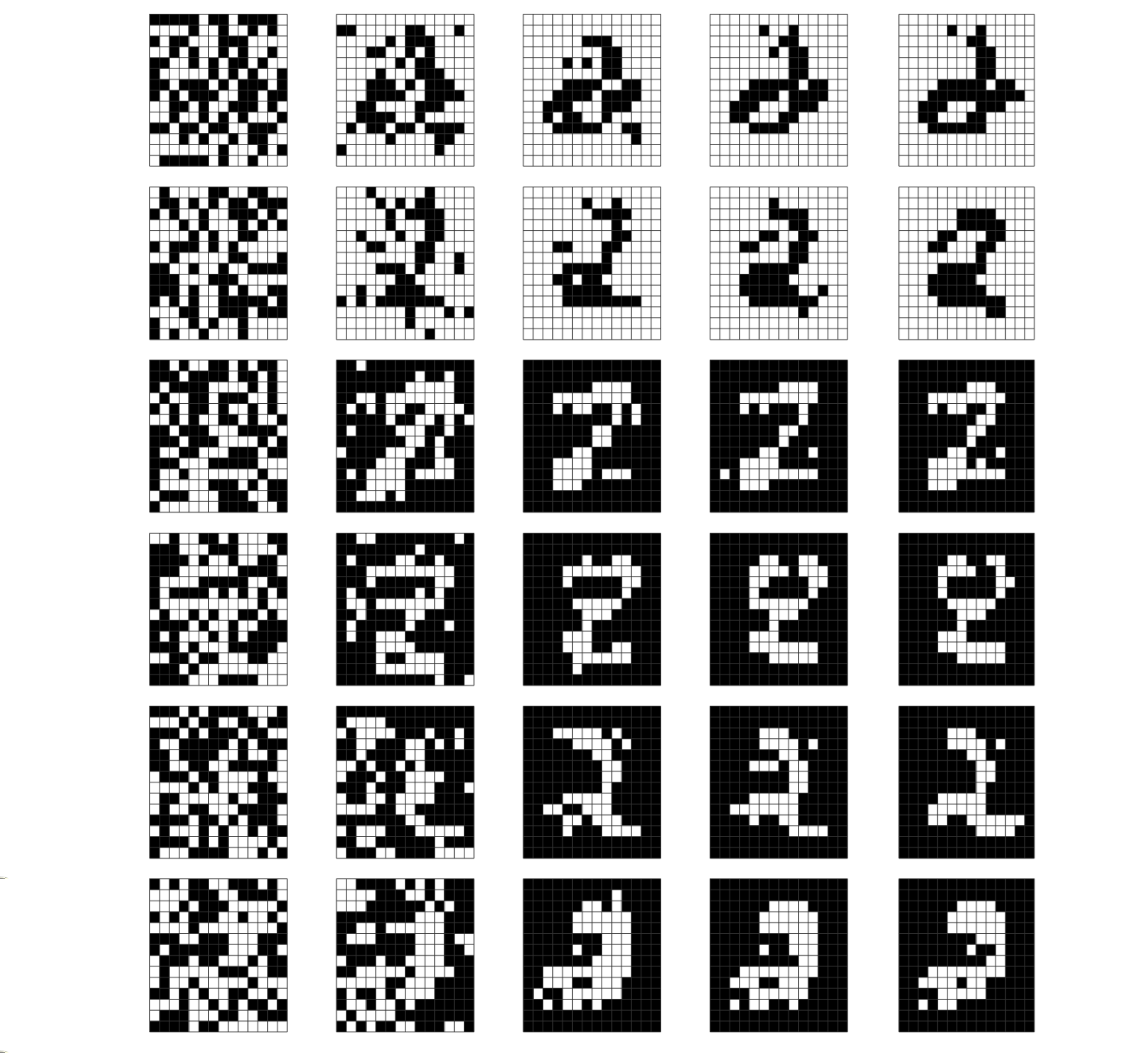}}
\caption{\label{fig:S6}The processes of generating six independent images for the digit `2'.}
\end{figure}
\begin{figure}[h]
\centerline{\includegraphics[width=3.9in]{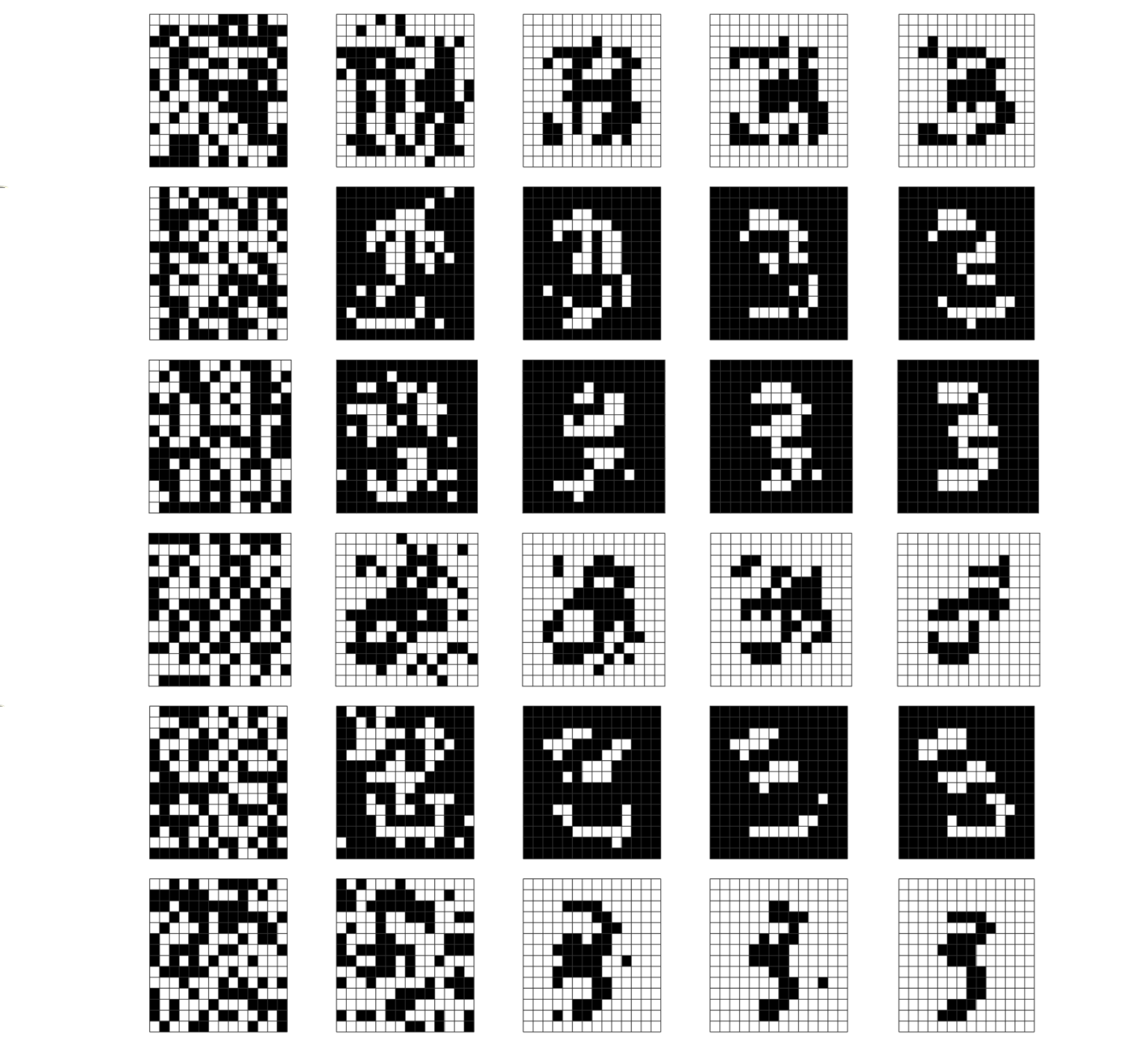}}
\caption{\label{fig:S7}The processes of generating six independent images for the digit `3'.}
\end{figure}
\begin{figure}[h]
\centerline{\includegraphics[width=3.9in]{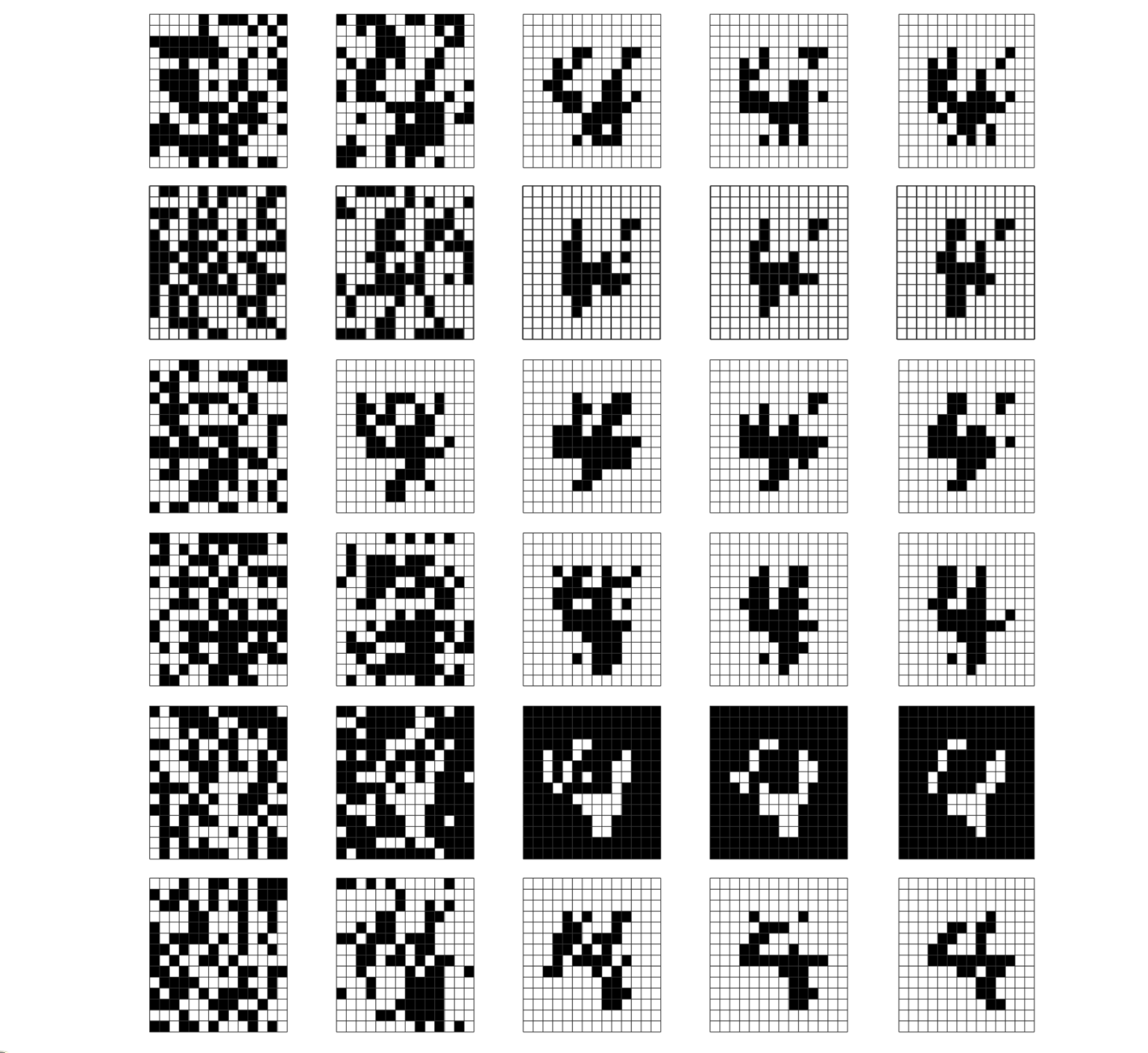}}
\caption{\label{fig:S8}The processes of generating six independent images for the digit `4'.}
\end{figure}
\begin{figure}[h]
\centerline{\includegraphics[width=3.9in]{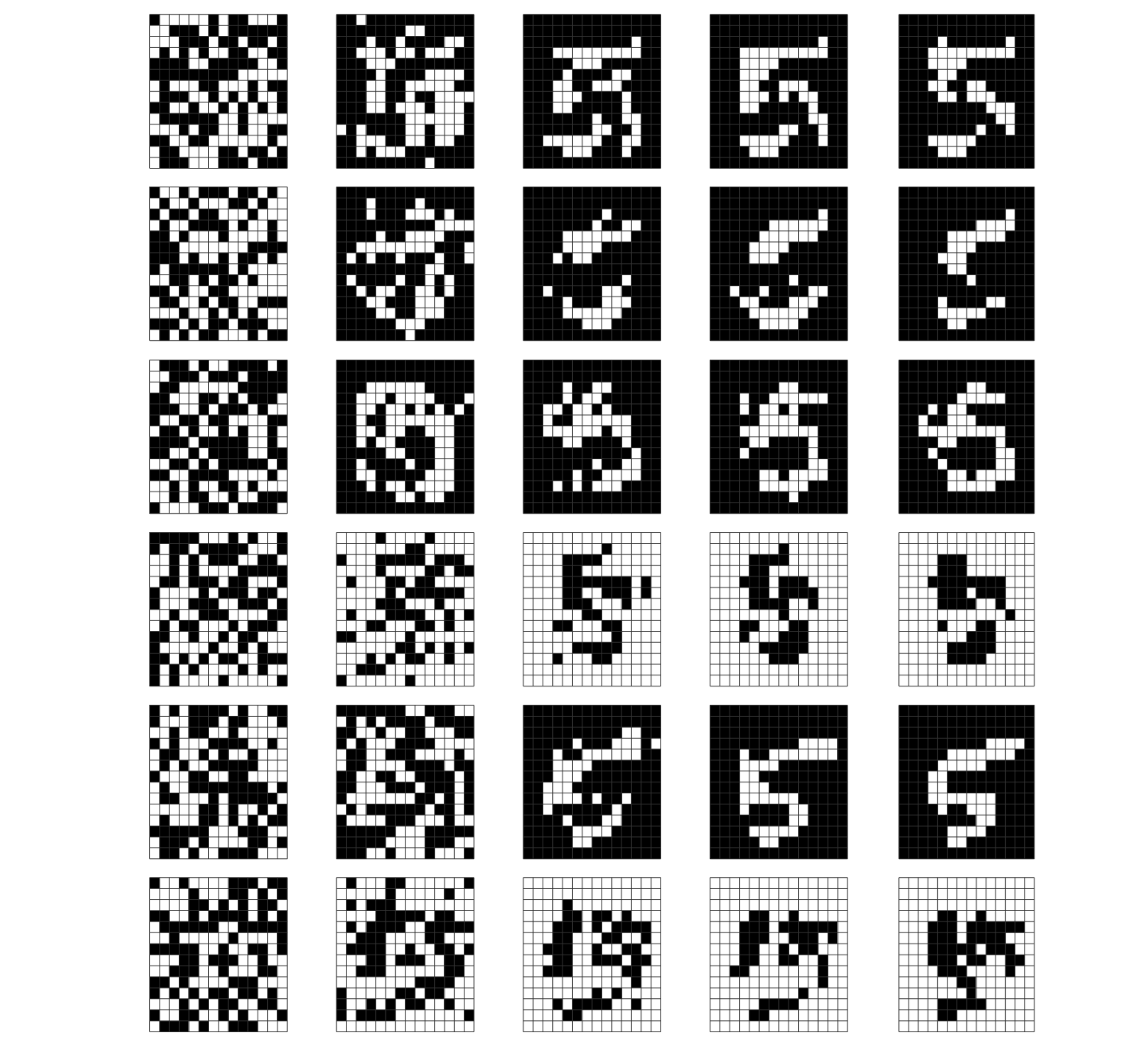}}
\caption{\label{fig:S9}The processes of generating six independent images for the digit `5'.}
\end{figure}
\begin{figure}[h]
\centerline{\includegraphics[width=3.9in]{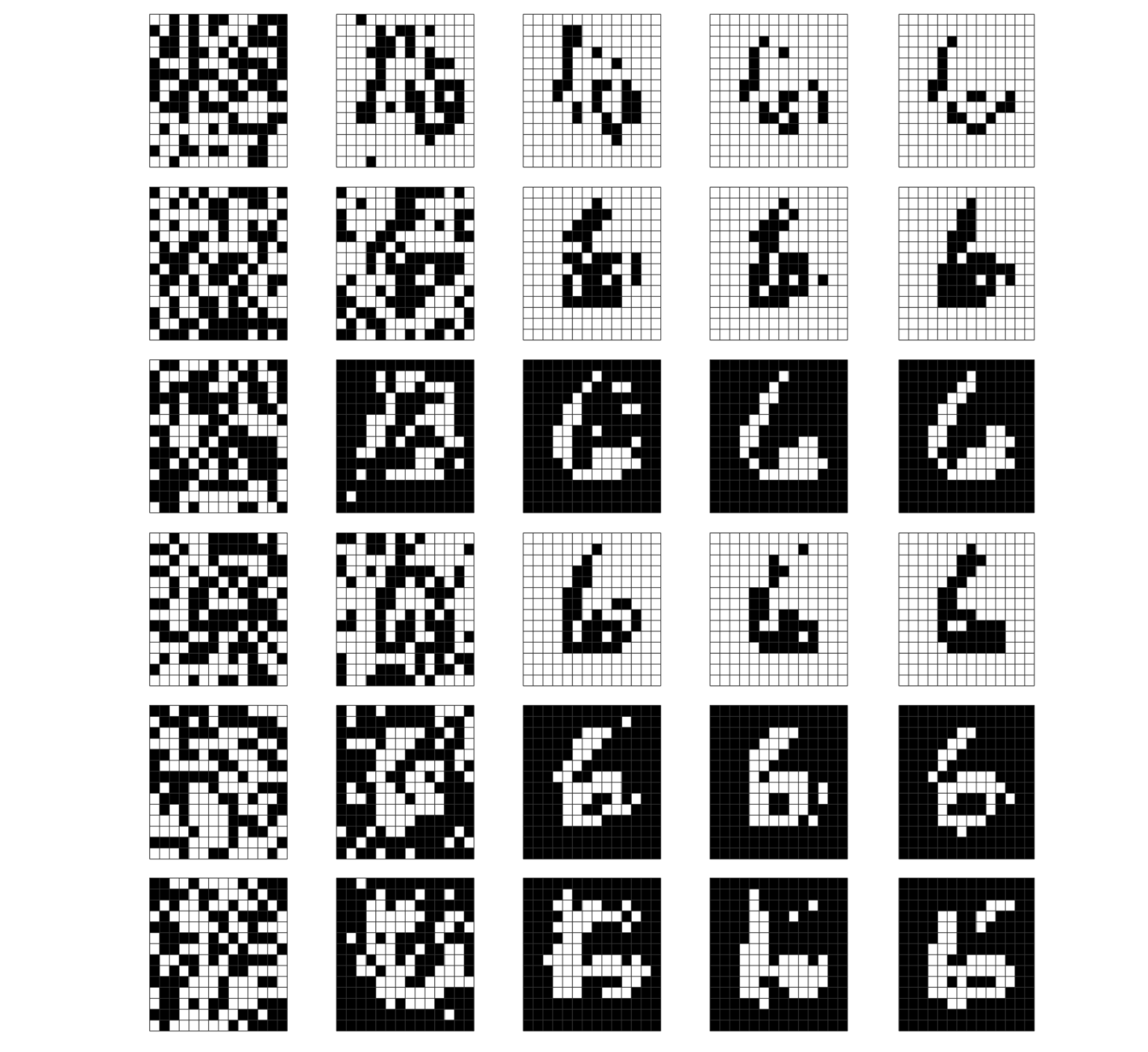}}
\caption{\label{fig:S10}The processes of generating six independent images for the digit `6'.}
\end{figure}
\begin{figure}[h]
\centerline{\includegraphics[width=3.9in]{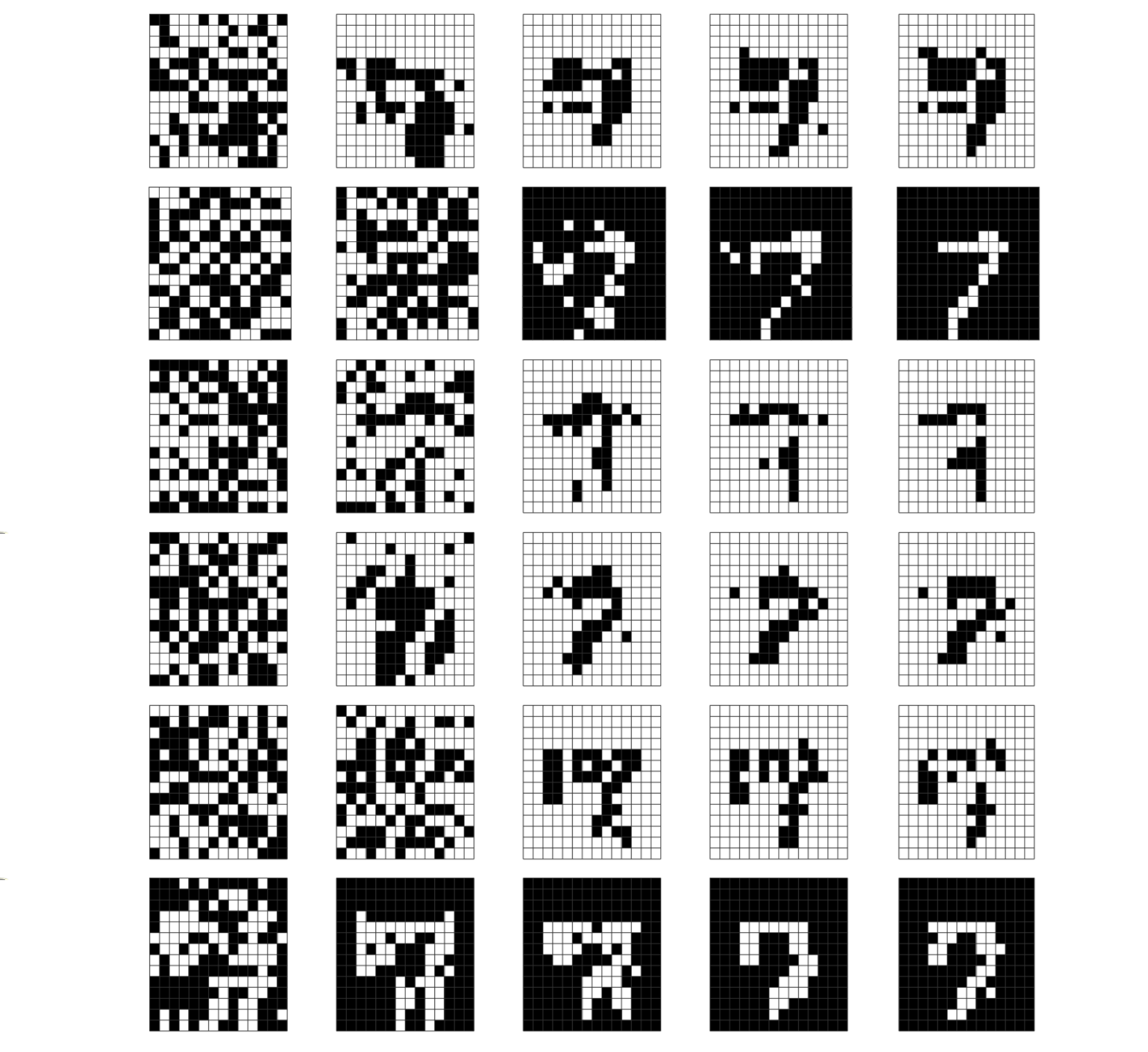}}
\caption{\label{fig:S11}The processes of generating six independent images for the digit `7'.}
\end{figure}
\begin{figure}[h]
\centerline{\includegraphics[width=3.9in]{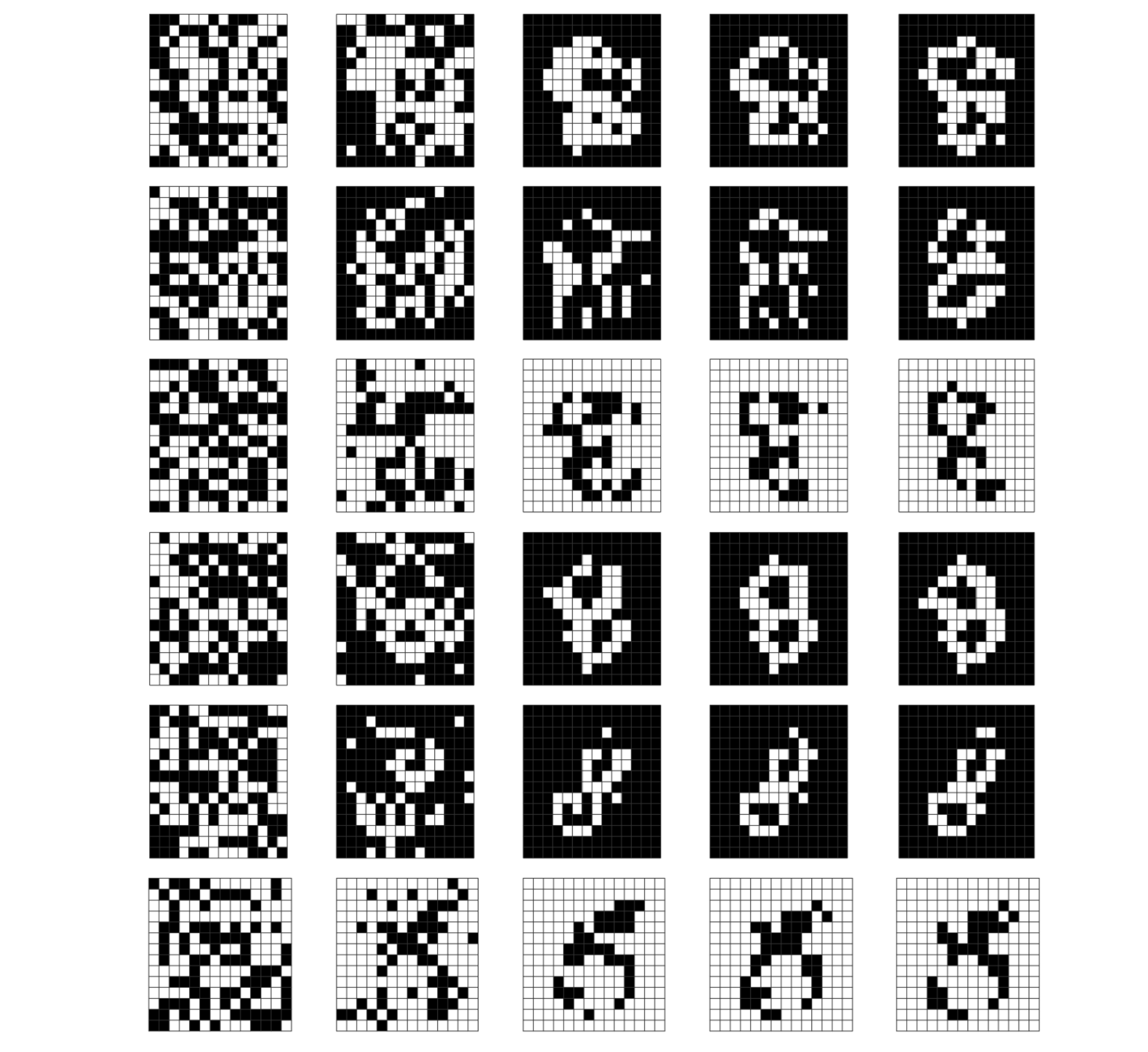}}
\caption{\label{fig:S12}The processes of generating six independent images for the digit `8'.}
\end{figure}
\begin{figure}[h]
\centerline{\includegraphics[width=3.9in]{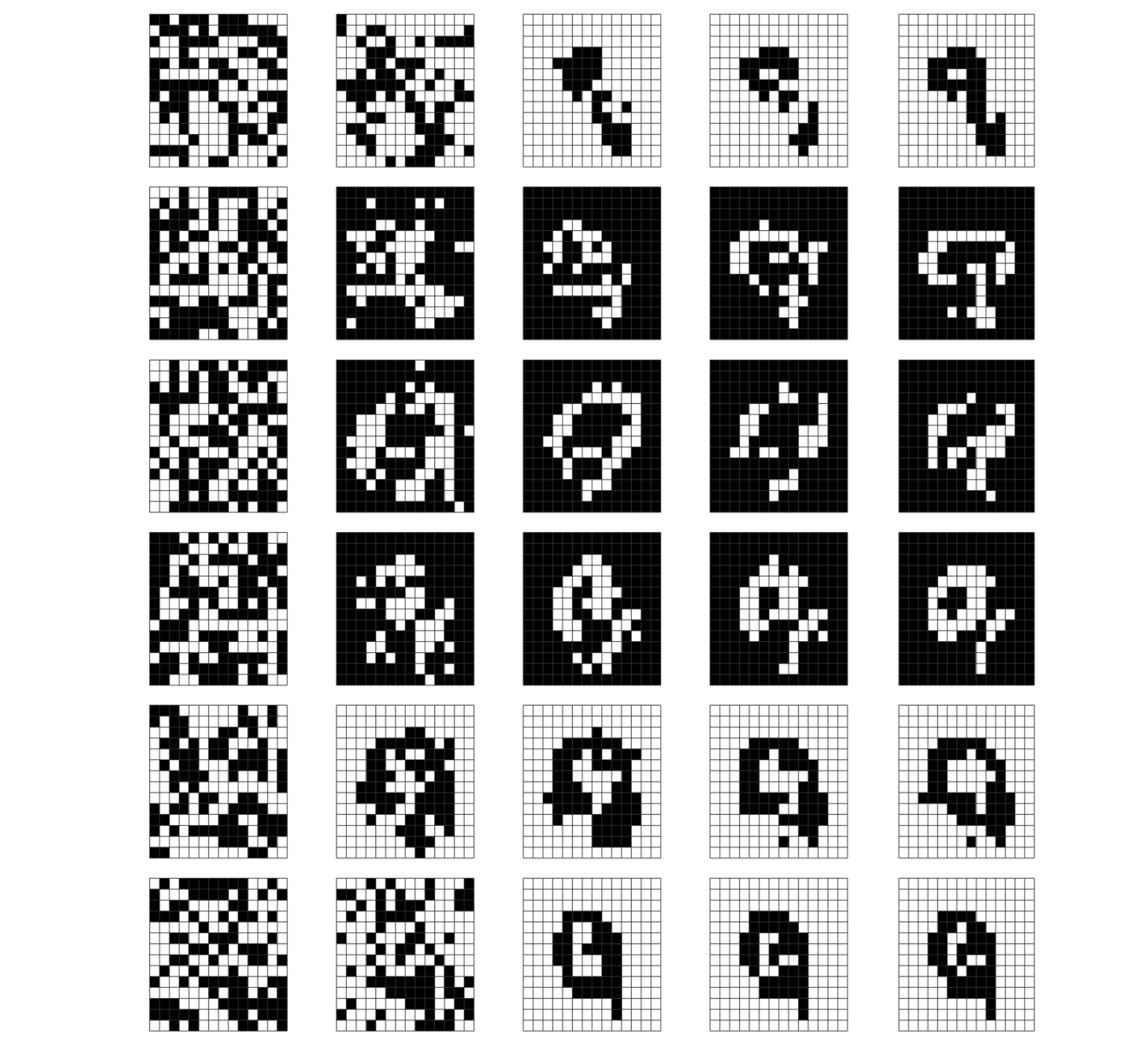}}
\caption{\label{fig:S13}The processes of generating six independent images for the digit `9'.}
\end{figure}
\begin{figure}[h]
\centerline{\includegraphics[width=3.7in]{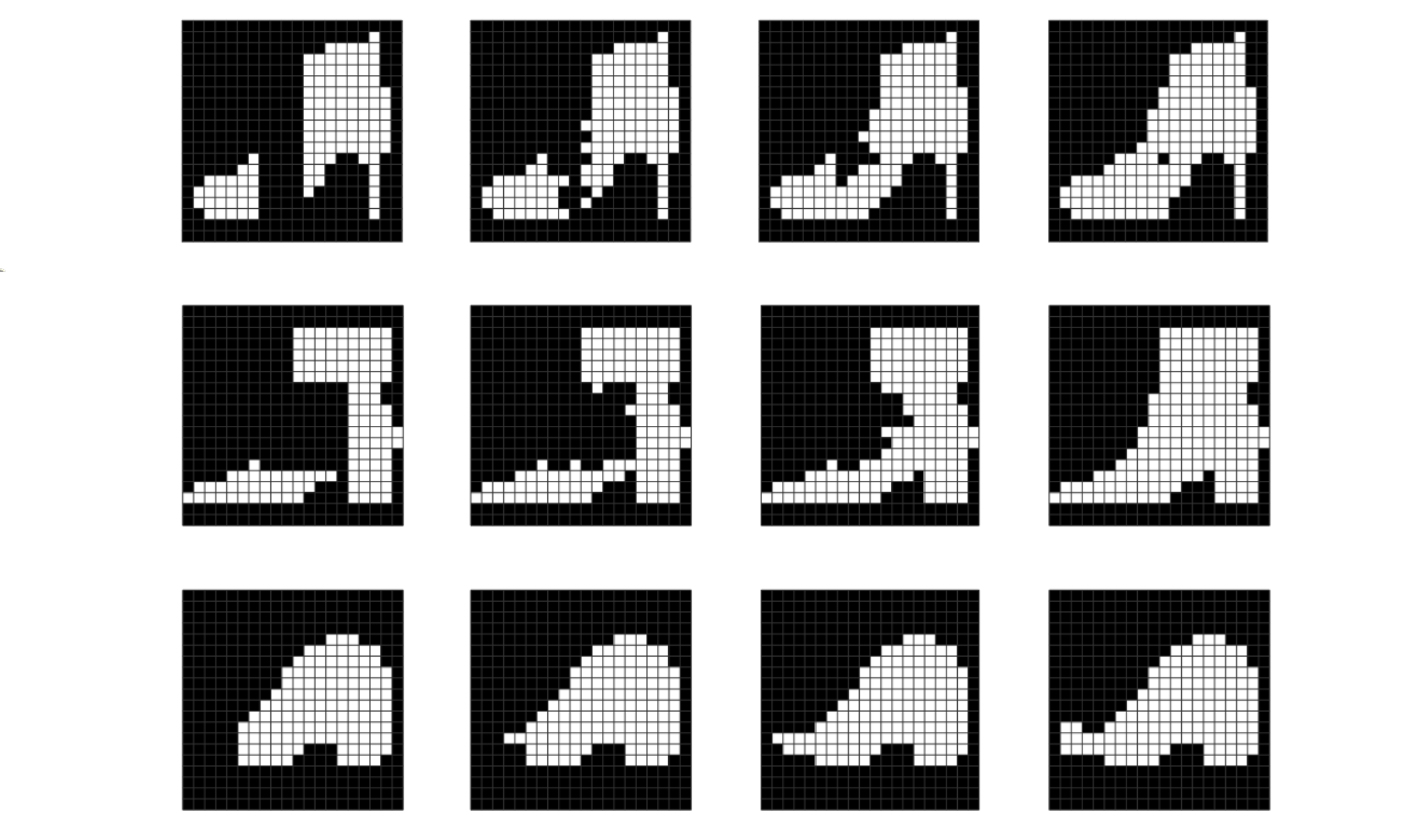}}
\caption{\label{fig:S14}The image restoration processes for occluded `Boot' images in test dataset.}
\end{figure}
\begin{figure}[h]
\centerline{\includegraphics[width=3.7in]{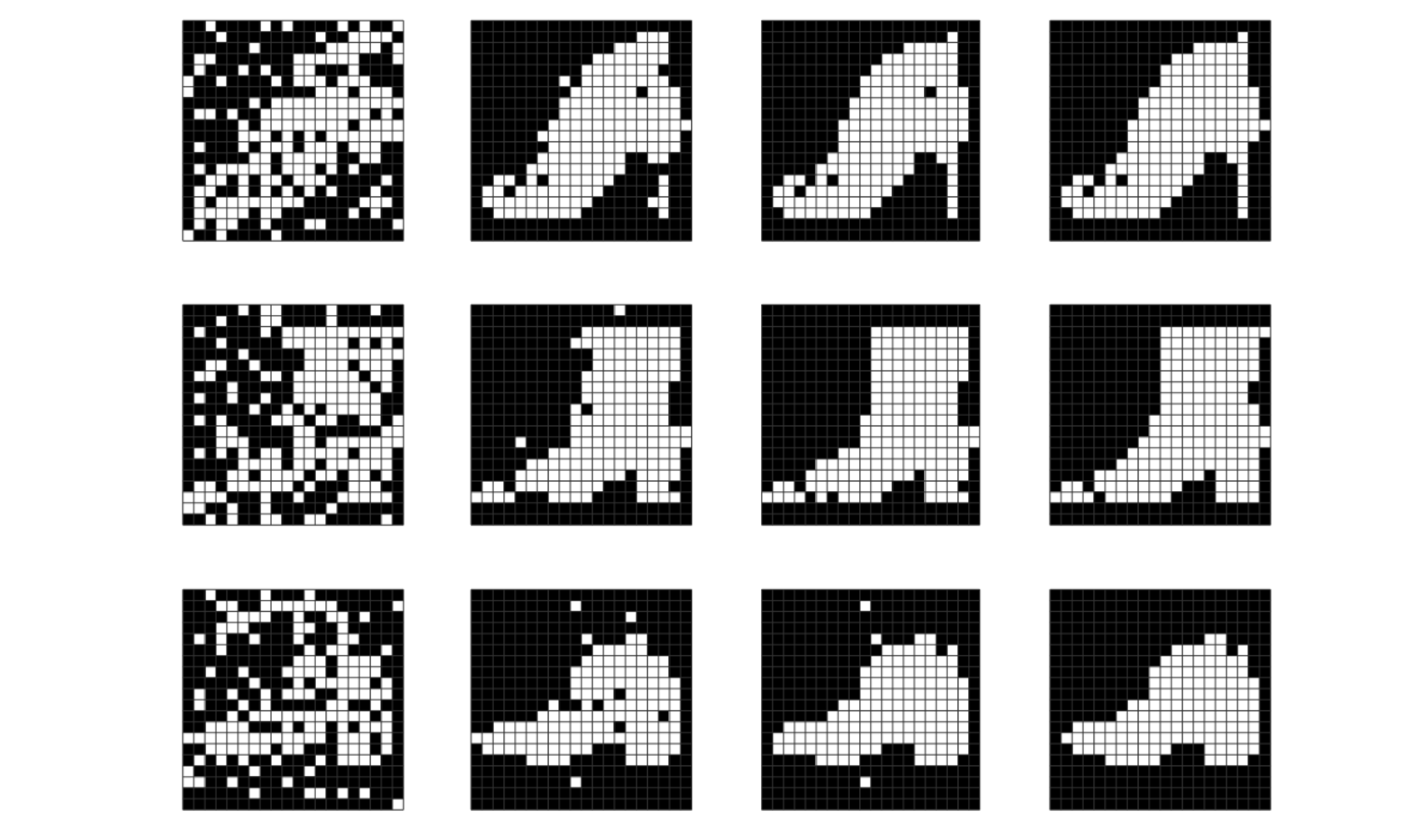}}
\caption{\label{fig:S15}The image restoration processes for `Boot' images with random noise in test dataset.}
\end{figure}

\begin{table}[h]
\centerline{\includegraphics[width=5in]{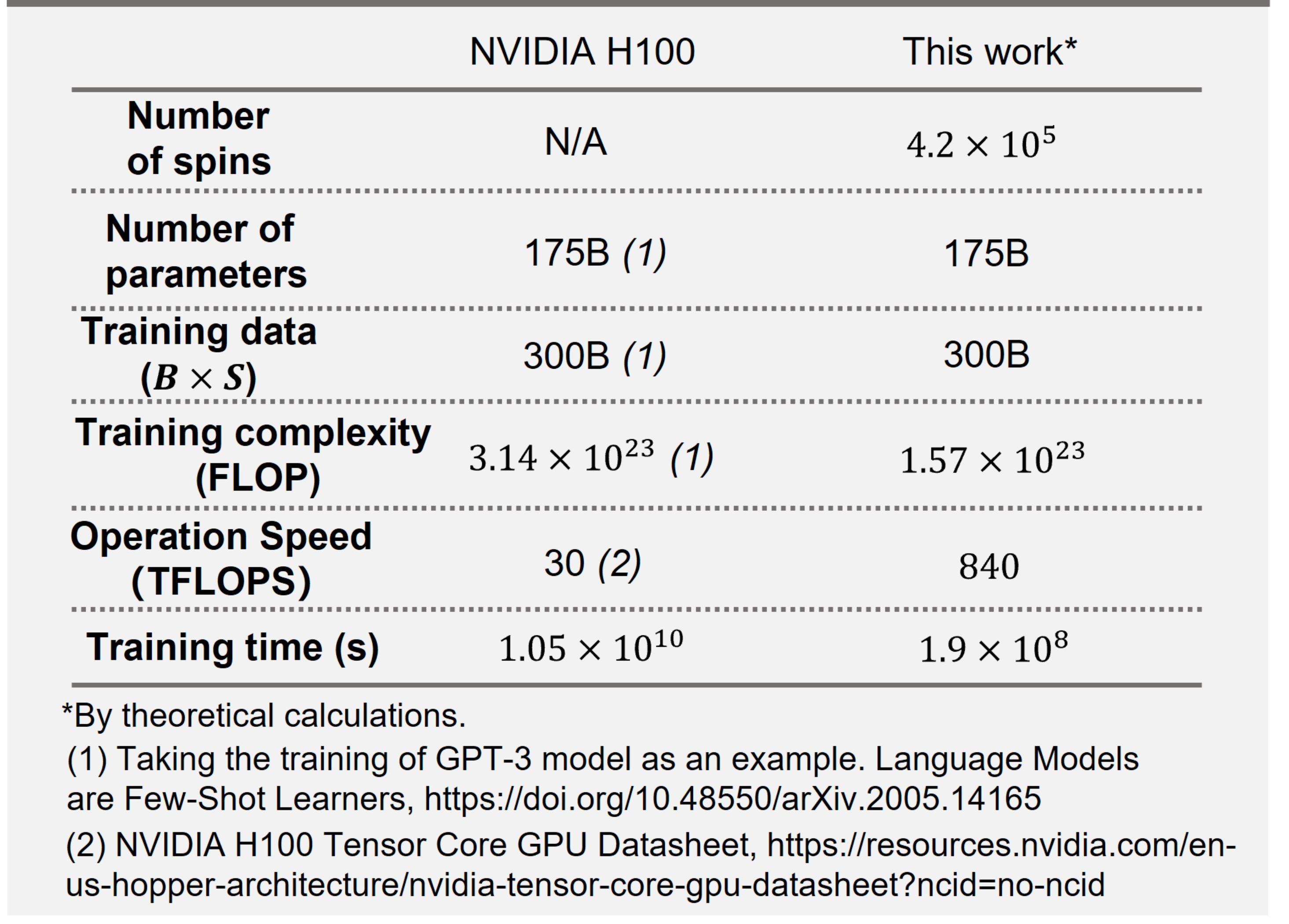}}
\caption{\label{Table:S1} Estimation of the training time by PRBM, in comparison with the NVIDIA H100, for the GPT-3 model with the same number of parameters and training data.}
\end{table}
\clearpage
\newpage

\end{document}